\documentclass[11pt]{article}
\usepackage{makeidx}   
\usepackage{graphicx}  
\usepackage{multicol}  
\makeindex             

\newtheorem{lemma}{Lemma}
\newtheorem{proposition}{Proposition}
\newtheorem{theorem}{Theorem}
\newtheorem{definition}{Definition}

\begin{document}

\author{Micha\l{} Horodecki, Pawe\l{} Horodecki and Ryszard Horodecki}

\title{Mixed-state entanglement and quantum communication}
\maketitle

We present basics of mixed-state entanglement theory. The first part  
of the  article is devoted  to mathematical characterizations of 
entangled  states. In second part we discuss the question of using 
mixed-state  entanglement for quantum communication. In particular, 
a type of  entanglement that is not directly useful for quantum 
communcation  (called bound entanglement) is  analysed in detail. 

This text is part of a book entitled {\it Quantum Information: An Introduction to 
Basic Theoretical Concepts and Experiments} by G. Alber, T. Beth,
M. Horodecki, P. Horodecki, R. Horodecki, M. R\"otteler, 
H. Weinfurter, R. Werner and  A. Zeilinger, published in {\it Springer 
Tracts in Modern Physics}, July 2001.

\tableofcontents


\pagenumbering{arabic}



\def\blacksquare{\vrule height 4pt width 3pt depth2pt}
\font\Bbb =msbm10  scaled \magstephalf
\def\Real{{\hbox{\Bbb R}}} \def\Complex{{\hbox {\Bbb C}}}
\def\id{{\hbox{\Bbb I}}}
\def\dag{\dagger}
\def\lcal{{\cal L}}
\def\hcal{{\cal H}}
\def\mcal{{\cal M}}
\def\fcal{{\cal F}}
\def\acal{{\cal A}}
\def\ra{\rangle}
\def\la{\langle}
\def\dim{\mathop{\rm dim}}
\def\trace{{\rm Tr}}
\def\be{\begin{equation}}
\def\ee{\end{equation}}
\def\ba{\begin{array}}
\def\ea{\end{array}}
\def\beq{\begin{eqnarray}}
\def\eeq{\end{eqnarray}}
\def\bei{\begin{itemize}}
\def\eei{\end{itemize}}

\part{Introduction}
\def\BE{bound entangled (BE)}
\def\FE{free entangled (FE)}
\def\UPB{basis!product!unextendible (UPB)}
Quantum entanglement\index{entanglement} is one
of the most striking features of quantum
formalism \cite{Peres-book}. It can be expressed as follows:
{\it
If two systems\index{system} interacted in
the past it is,  in general, not possible to
assign a single state vector\index{vector!state} to either of the to
subsystems\index{subsystem}%
} \cite{desp}. This is what is sometimes called the principle of
non-separability\index{principle!non-separability}. A common
example of entangled state\index{state!entangled}
is the singlet state\index{state!singlet}
\cite{Bohm}
\be
\psi_-={1\over \sqrt2}(|01\ra-|10\ra)\;.
\label{singlet}
\ee
One can see that it cannot be represented as a product of individual
vectors\index{vector} describing states\index{state} of
subsystems\index{subsystem}.
Historically, entanglement\index{entanglement} was first
recognised by Einstein, Podolsky and
Rosen
(EPR) \cite{EPR} and by Schr\"odinger \cite{Schrodinger}\footnote{
In fact, entangled quantum states\index{state!quantum}\index{quantum!state} have been
used in investigations
of the properties of atomic and molecular
systems\index{system} \cite{Heitler}}.
In their famous
paper EPR
suggested a description of the
world\index{world} (called ``local realism'') which assigns
an independent and objective reality to the physical properties of the well
separated subsystems\index{subsystem} of a compound
system\index{system!compound}.
Then EPR applied the criterion of local realism to  predictions 
associated with an entangled state \index{state!entangled}
to conclude that
quantum mechanics\index{quantum!mechanics}
is incomplete.
EPR criticism was the source of many discussions concerning fundamental
differences between quantum and classical description of nature.

The most significant progress toward the resolution of the EPR problem was
made by Bell \cite{Bell} who proved
that the local realism implies constraints
on the predictions of spin\index{spin!correlations} correlations in
the form of inequalities
(called Bell's inequalities)\index{inequalities!Bell} which can be
{\it violated} by  quantum mechanical
predictions\index{quantum!predictions} for the system\index{system} in 
the  state \index{state} (\ref{singlet}).
The latter feature of quantum mechanics\index{quantum!mechanics} called
usually {\it nonlocality}\index{nonlocality}
is one of the most apparent manifestations of
quantum entanglement\index{entanglement}.

Information theoretic aspect of entanglement\index{entanglement} was
first considered by
Schr\"odinger who wrote in the context of the EPR problem:
``Thus one disposes provisionally (until
the entanglement\index{entanglement} is resolved by actual
observation) of only a {\it common} description of the two in that
space of
higher dimension. This is the reason that knowledge of the
individual systems\index{system}
can {\it decline} to the scantiest, even to zero, while that of the combined
system\index{system} remains continually maximal. Best possible knowledge of
a whole
does {\it not} include best possible knowledge of its parts -- and that is
what keeps coming back to haunt us'' \cite{Schrodinger}.
In this way Schr\"odinger
recognised a profoundly non-classical relation between the
information\index{information}
which an entangled state\index{state!entangled} gives us
about the whole system\index{system} and the
information\index{information}
which it gives us about the subsystems\index{subsystem}.

The recent  development of quantum
information\index{information!quantum}\index{quantum!information} theory showed
that entanglement\index{entanglement} can have important {\it practical} applications
(see e.g. \cite{world}).
In particular  it turned out that 
entanglement\index{entanglement} can be used as a {\it resource}
for communication of quantum\index{quantum!state}
states\index{state!quantum}\index{communication!of quantum states} in
astonishing process called
{\it quantum teleportation}\index{quantum!teleportation}
\index{teleportation}
\cite{Bennett_tel}. In  the latter
a quantum
state\index{state!quantum}\index{quantum!state} is transmitted
by use of a pair of particles in singlet state\index{state!singlet}
(\ref{singlet})
shared by the sender and receiver (typically Alice and Bob), and two
bits \index{bit}
of classical communication\index{communication!classical}.
However, in real conditions, due to
interaction\index{interaction with environment} with
environment\index{environment}, called
decoherence\index{decoherence},  we
encounter mixed states\index{state!mixed} rather than
pure ones. They can still possess
some residual entanglement\index{entanglement}. More specifically, a mixed
state\index{state!mixed} is considered to be entangled if it is not a
mixture\index{mixture!of product states} of product
states\index{state!product}
\cite{Werner}. In mixed states\index{state!mixed} the quantum
correlations\index{quantum!correlations} are weakened, hence
the manifestations of mixed-state
entanglement\index{entanglement!mixed-state} can be very subtle
\cite{Werner,Popescu94,hidden}. Nevertheless, it appears that it can be
used as a resource  for
quantum
communication\index{quantum!communication}\index{communication!quantum}.
Such 
possibility is due to discovery of {\it distillation}\index{distillation}
of entanglement\index{entanglement!distillation} \cite{Bennett_pur}: by
manipulation over noisy pairs,
involving local operations\index{operations!local} and classical
communication\index{communication!classical},
Alice and Bob can obtain singlet pairs, and apply
teleportation\index{teleportation}.
This procedure  provides a  powerful protection
of the quantum data\index{quantum!data}
transmission\index{transmission} against
environment\index{environment}.

Consequently, the fundamental problem was to investigate the structure
of mixed-state entanglement\index{entanglement!mixed-state}, especially
in the context of quantum
communication\index{communication!quantum}\index{quantum!communication}.
These investigations have lead to discovery
of {\it discontinuity} in the structure of mixed-state
entanglement\index{entanglement!mixed-state}.
It appeared that there are at least two qualitatively different types
of entanglement\index{entanglement} \cite{bound}:
{\it free}\index{entanglement!free} -- useful
for quantum
communication\index{communication!quantum}\index{quantum!communication},
and {\it bound} - a
non-distillable\index{entanglement!bound}\index{entanglement!non-distillable},
very weak and mysterious
type of entanglement\index{entanglement!types}.

The present contribution is divided into two main parts. In the first one
we report results of investigation of {\it mathematical} structure of
entanglement\index{entanglement}. The main question is: given a
mixed state\index{state!mixed}, is it entangled
or not? We present powerful tools that allow  to obtain the answer
in many interesting cases. Crucial role is here played by the connection
between entanglement\index{entanglement} and
theory of {\it positive maps}\index{map!positive}
\cite{sep}.
In contrast
to {\it completely positive maps}\index{map!completely positive}
\cite{Kraus}, the positive maps\index{map!positive}
were not applied in physics so far.
The second part is devoted to application of
the entanglement of mixed
states\index{entanglement!mixed-state}\index{state!mixed} to
quantum
communication\index{communication!quantum}\index{quantum!communication}. Now,
the leading question is: given entangled
state\index{state!entangled}, can it be distilled?
The mathematical tools worked out in the first part allowed to answer the
question. Surprisingly, the answer did not simplify the picture,
but rather revealed a new horizon including the basic question:
what is the role of bound entanglement\index{entanglement!free}
in Nature?

Since entanglement\index{entanglement} is a basic
ingredient of the quantum
information\index{quantum!information}\index{quantum!information} theory,
the scope of application of the presented research
goes far beyond the quantum
communication\index{quantum!communication}\index{communication!quantum}
problem. The insight into structure of entanglement of
mixed states\index{state!mixed}\index{entanglement!mixed-state} can be
helpful in many subfields
of quantum information\index{quantum!information}\index{information!quantum}
theory, including  quantum
computing\index{quantum!computing}, quantum
cryptography\index{quantum!cryptography} etc.

Finally, it must be emphasised that
our approach will be basically {\it qualitative}. Thus
we will not review here the beautiful work performed in the domain of
{\it quantifying} 
entanglement\index{entanglement!quantifying}
\cite{huge,Plenio,VP98,Tarrach,Vidal} (we will only touch this
subject in the second part). Due to limited volume of the present
contribution, we will also restrict considerations to 
entanglement\index{entanglement!of bipartite systems}
of {\it bipartite} systems\index{system!bipartite}, even though
a number of results
has been recently obtained for {\it multipartite}
systems\index{system!multipartite}
(see e.g. \cite{Murao,UPB1}).

\part{Entanglement of mixed states: characterisation}
We will deal with the states\index{state}  on the
finite dimensional Hilbert
space\index{space!Hilbert} ${\cal H}_{AB}={\cal H}_A\otimes {\cal H}_B$.
A system\index{system} described by Hilbert
space\index{space!Hilbert} $\hcal_{AB}$ we will call $n\otimes m$
system\index{system}, where $n$ and $m$ are dimensions of the
spaces\index{space} $\hcal_A$ and $\hcal_B$
respectively.
An operator\index{operator} $\varrho$ acting on $\cal H$ is a
state\index{state} if $\trace\varrho=1$
and if it is a positive operator\index{operator!positive} i.e.
\begin{equation}
\trace \varrho P\geq0
\end{equation}
for any projectors\index{projector} $P$ (equivalently, positivity of
operator\index{operator} means that it is
Hermitian \index{operator!Hermitian} and has nonnegative 
eigenvalues\index{eigenvalues}).

A state\index{state} acting on Hilbert space\index{space!Hilbert}
${\cal H}_{AB}$
is called separable\footnote{The
presented definition of separable states\index{state!separable} is due to Werner \cite{Werner} who
called them classically correlated states\index{state!classically correlated}.}
if it can be approximated
in the trace norm\index{norm!trace} by the
states\index{state} of the form
\begin{equation}
\varrho=\sum_{i=1}^kp_i\varrho_i\otimes\tilde \varrho_i\;,
\label{sep}
\end{equation}
where $\varrho_i$ and $\tilde\varrho_i$ are states\index{state} on ${\cal H}_A$ and
${\cal H}_B$ respectively.
In finite dimensions one can use simpler definition \cite{transp} (see 
also \cite{VP98}):
$\varrho$ is separable if it is of the form (\ref{sep}) for some $k$
(one can always find $k\leq\dim \hcal_{AB}^2$). Note that the property
of being  entangled or not does not change if one subjects  
the state\index{state} to a
product unitary
transformation\index{transformation!unitary!product} $\varrho\rightarrow \varrho'=U_1\otimes U_2\varrho U_1^\dagger
\otimes
U_2^\dagger$. The states\index{state} $\varrho$ and $\varrho'$ we call {\it equivalent}.

We will further need the following {\it maximally} entangled pure state
\index{state!entangled!maximally}\index{state!pure} of
$d\otimes d$ system\index{system!$d\otimes d$}
\be
\psi_+^d={1 \over \sqrt d}\sum_{i=1}^d|i\ra\otimes|i\ra\;.
\ee
The corresponding  projector\index{projector} we will denote by $P_+^d$ (the superscript $d$
will be usually omitted).
Then for any state\index{state} $\varrho$ the quantity $F=\la\psi_+| \varrho|\psi_+ \ra$
is called {\it singlet fraction}\index{fraction!singlet}%
\footnote{In fact, the state\index{state} $\psi_+$
used in definition of singlet fraction\index{fraction!singlet} is a
local transformation\index{transformation!local} of
true singlet state\index{state!singlet}. Nevertheless, we will keep the
name ``singlet fraction''\index{fraction!singlet}
while using the state\index{state} $\psi_+$ being more
convenient from technical reasons.}.
In general, by maximally entangled states\index{state!entangled!maximally}
we will mean vectors\index{vector} $\psi$ that are
 equivalent to $\psi_+$
\[
\psi=U_1\otimes U_2 \psi_+
\]
where $U_1,U_2$ are unitary transformations\index{transformation!unitary}.
The most common two-qubit\index{qubit}
maximally entangled state\index{state!entangled!maximally} is
the {\it singlet} state\index{state!singlet} (\ref{singlet}). One can
define {\it fully entangled fraction}\index{fraction!fully entangled} of
a state\index{state} $\varrho$ of $d\otimes d$
system\index{system!$d\otimes d$} by
\be
\fcal(\varrho)=\max_\psi \la\psi|\varrho|\psi\ra\;,
\ee
where the maximum is taken over all maximally entangled
vectors\index{vector!entangled!maximally} of $d\otimes d$ 
system\index{system!$d\otimes d$}.

\section{Pure states}
If $\varrho$ is a {\it pure} state\index{state!pure} i.e. $\varrho=|\psi\ra\la\psi|$, then it is
easy to check if it is entangled
or not. Indeed, the above definition implies that it is separable if and only
if $\psi=\psi_A\otimes \psi_B$, i.e. if either of its  reduced density
matrices is pure state\index{state!pure}. Thus it suffices to find
eigenvalues\index{eigenvalues}  of either of the
reductions. Equivalently one can refer to Schmidt decomposition%
\index{decomposition!Schmidt} \cite{Schmidt}
of the state\index{state}. As one knows, for any pure
state\index{state!pure} $\psi$
there exist bases\index{basis} $\{e^A_i\}$,
$\{e^B_i\}$ in spaces\index{space} $\hcal_A$ and $\hcal_B$ such that
\be
\psi=\sum_{i=1}^ka_i |e^A_i\ra\otimes | e^B_i\ra,\quad k\leq\dim\hcal_{AB}
\ee
with  positive coefficients $a_i$  are called Schmidt coefficients.
Then the state\index{state} is entangled if at least two coefficients do not vanish.
One finds that the positive eigenvalues\index{eigenvalues} of either
of the reductions are
equal to squares of the Schmidt coefficients.
In next section we will introduce a series of necessary conditions for
separability \index{condition!for separability!necessary} for
mixed states\index{state!mixed}.
It turns out that all of them
are equivalent to separability in the case of pure
states\index{state!pure} \cite{Gisin_Bell,renyi}.

\section{Some necessary conditions for separability of mixed states}

A condition that is satisfied by separable
states\index{state!separable} will be called
{\it separability criterion}\index{criterion!separability}. If a
separability criterion\index{criterion!separability} is violated by
state\index{state}, the state must be entangled\index{state!entangled}.
It is important to have strong
separability criteria\index{criterion!separability}, i.e. the ones
that are violated by possibly the
largest number of states\index{state}.

Since violation of Bell inequalities\index{inequalities!Bell} is a
manifestation of quantum
entanglement\index{entanglement}, a natural separability
criterion\index{criterion!separability}
is
constituted by Bell inequalities\index{inequalities!Bell}. 
In \cite{Werner} Werner first pointed out that  separable
states\index{state!separable} must satisfy all possible
Bell inequalities\index{inequalities!Bell}\footnote{In \cite{Werner}
Werner also provided a very useful criterion based on so-called {\it flip}
operator (see Sect. \ref{examples}).}.
The common Bell inequalities\index{inequalities!Bell} derived by
Clauser, Horne, Shimony and Holt (CHSH) are given by \cite{CHSH}
\be
\trace \varrho {\cal B}\leq 2\;,
\ee
where the Bell-CHSH observable\index{observable!Bell-CHSH}
$\cal B$ is given by
\be
{\cal B}=\vec{\hat a}\vec{\sigma}\otimes(\vec{\hat b}+\vec{\hat b'})
\vec{\sigma}+\vec{\hat a'}\vec{\sigma}\otimes
(\vec{\hat b}-\vec{\hat b'})\vec{\sigma}\;
\ee
where $\vec{\hat a},\vec{\hat a'},\vec{\hat b},\vec{\hat b'}$ are 
arbitrary unit
vectors\index{vector!unit}
in $\Real^3$, $\vec{\hat a}{\bf \sigma}=\sum_{i=1}^3a_i\sigma_i$,
and $\sigma_i$ are Pauli matrices.  For any given set of the
vectors\index{vector} we have a different inequality.  In \cite{bell} one derived the
condition for a two-qubit\index{qubit}%
\footnote{A {\it qubit}\index{qubit} is the elementary unit of quantum
information\index{information!quantum}\index{quantum!information} and
denotes two
level quantum system\index{quantum!system}\index{system!quantum} (i.e. 
$2\otimes 2$ system\index{system!$2\otimes2$}) \cite{Schumacher}.}
state\index{state} equivalent to satisfying all
the inequalities jointly. It has the following form
\be
M(\varrho)\leq 1\;,
\label{mbell}
\ee
where $M$ is constructed in the following way. One considers the $3\times 3$
real matrix  $T$ with entries $T_{ij}\equiv \trace \varrho\sigma_i\otimes \sigma_j$.
Then $M$ is equal to the sum of two greater eigenvalues\index{eigenvalues}
of the matrix $T^\dagger T$. This condition, characterises states\index{state}
violating the most common, and so far the strongest Bell
inequality\index{inequalities!Bell} for
two qubits\index{qubit} (see \cite{garg} in this context).
Being interesting from the point of view of
nonlocality\index{nonlocality}, it appears
to be not a very strong separability criterion \index{criterion!separability}.
Indeed, there exists \cite{Werner} a large class
of entangled states\index{state!entangled} that satisfy all
standard Bell inequalities%
\index{inequalities!Bell}%
\footnote{See \cite{Popescu94,hidden,Marek} in the context of
more sophisticated nonlocality\index{nonlocality} criteria.}.

Another approach originated from Schr\"odinger \cite{Schrodinger}
observation that an entangled state\index{state!entangled} gives us more information about
the total system\index{system} than about subsystems\index{subsystem}. This
gave rise to a
series of entropic inequalities\index{inequalities!entropic} of
the form \cite{red,renyi}
\be
S(\varrho_A)\leq S(\varrho),\quad S(\varrho_B)\leq S(\varrho)\;,
\label{entrop-ineq}
\ee
where $\varrho_A=\trace_B\varrho$ and similarly for $\varrho_B$.
The above inequalities were proven 
\cite{red,renyi,xor,PSTT} to be satisfied by separable
states\index{state!separable}
for four different entropies\index{entropy} being particular
cases of Ren\'yi quantum
entropies\index{entropy!Ren\'yi}\index{quantum!entropies}
$S_\alpha=(1-\alpha)^{-1}\log\trace \varrho^\alpha$
\beq
&S_0=\log R(\varrho) &\quad   \label{rank}\\
&S_1=-\trace \varrho\log\varrho &\quad  \\
&S_2=-\log\trace\varrho^2  &	 \quad	\\
&S_{\infty}=-\log||\varrho||\;, &      \quad
\eeq
where $R(\varrho)$ denotes the rank\index{rank} of the
state\index{state} $\varrho$ (number of
non-vanishing eigenvalues\index{eigenvalues}).
The above inequalities are useful tools in many cases (as we will see
in Sect. \ref{sec-search} one of them allows to obtain bound on the possible
rank\index{rank} of the {\it bound entangled}
states\index{state!bound entangled (BE)}, still however
they are not very strong criteria.

A different approach, presented in Ref. \cite{huge}, is based on local
manipulations of entanglement\index{entanglement!manipulation}
(the approach was anticipated in Ref. \cite{hidden}).
The main line is of the following sort: a given
state\index{state} is entangled,
because the parties sharing many systems\index{system} (pairs of particles) in this
state\index{state}
can produce less number of pairs in highly entangled
state\index{state!entangled} (of easily ``detectable''
entanglement\index{entanglement}) by
local operations\index{operations!local} and classical
communication\index{communication!classical} (LQCC).
This approach initiated new field in quantum\index{quantum!information}
information\index{information!quantum} theory: manipulating
entanglement\index{entanglement!manipulation}.  The second part of this
contribution will be devoted to this field. It also initiated  the subject of
quantification of entanglement\index{entanglement!quantifying}. Still,
however, the seemingly simple
qualitative question of whether a given state\index{state} is entangled or not was not
solved.

A breakthrough was done by Peres \cite{Peres} who derived a surprisingly
simple but very strong criterion. He noted that a separable
state\index{state!separable}
remains positive operator\index{operator!positive} if subjected to
partial transposition (PT).
We will call it positive partial transposition (PPT) criterion%
\index{criterion!PPT}.

To define partial transposition, we will use matrix
elements\index{matrix!elements} of
a state\index{state} in some product basis \index{basis!product}:
\begin{eqnarray}
\varrho_{m \mu, n \nu}=\langle m| \otimes \la \mu | \,\varrho\,|
n \ra\otimes |\nu\ra\;,
\end{eqnarray}
where the kets with Latin (Greek) letters form
orthonormal basis \index{basis!orthonormal} in Hilbert
space\index{space!Hilbert} describing
first (second) system\index{system}. Hence the partial transposition of $\varrho$ is defined as:
\begin{eqnarray}
\varrho^{T_B}_{m \mu, n \nu}\equiv \varrho_{m \nu, n \mu}\;.
\end{eqnarray}
The form of the operator\index{operator} $\varrho^{T_B}$ depends on the
choice of basis\index{basis},  but its eigenvalues\index{eigenvalues} do not.
We will  say that a state\index{state!PPT} is PPT if $\varrho^{T_B}\geq0$; otherwise
we will say that the state\index{state!NPT} is NPT.
The partial transposition is easy to perform in matrix notation.
Since the state\index{state} of $m \otimes n$
system\index{system!$m\otimes n$} can be written as
\begin{eqnarray}
\varrho=\left[ \begin{array}{ccc}
         A_{11} \ ... \  A_{1m} \\
	 ... \ ... \ ... \\
         A_{m1} \ ... \ A_{mm} \\
       \end{array}
      \right ]
\end{eqnarray}
with $n\times n$ matrices
$A_{i j}$  acting on the second ($\Complex^n$) space\index{space}.
They are defined
by their matrix elements\index{matrix!elements} as
$\{ A_{i j} \}_{\mu \nu} \equiv  \varrho_{i \nu, j \mu}$.
Then the partial transposition will be realised simply by transposition
(denoted by $T$) of all of these matrices, namely:
\begin{eqnarray}
\varrho^{T_B}=\left[ \begin{array}{ccc}
         A_{11}^T \ ... \  A_{1m}^T \\
	 ... \ ... \ ... \\
         A_{m1}^T \ ... \ A_{mm}^T \\
       \end{array}
      \right ]\;.
\label{PT-wykonanie}
\end{eqnarray}

Now \cite{Peres} for any separable state\index{state!separable} $\varrho$, the
operator\index{operator}
$\varrho^{T_B}$ must have still nonnegative eigenvalues\index{eigenvalues}.
Indeed, consider partially transposed separable
state\index{state!separable}:
\be
\varrho^{T_B}= \sum_ip_i\varrho_i\otimes(\tilde\varrho_i)^T\;.
\ee
Since the state\index{state} $\tilde\varrho_i$ remains positive under transposition,
so does the total state\index{state}.

Note that what distinguishes the Peres criterion\index{criterion!Peres}
 from the earlier ones is
that it is {\it structural}. In other words, it does not say that some scalar
function of a state\index{state} satisfies some inequality, but
it imposes constraints on
the structure of the
operator\index{operator} resulting from  PT. Thus the criterion
amounts to satisfying of many inequalities at the same time.
In next section  we will see that there is also another crucial feature
of the criterion: it involves transposition that is positive map%
\index{map!positive} but
is {\it not} completely positive one.  This feature abstracted from the
Peres criterion\index{criterion!Peres} allowed to find intimate connection
between entanglement\index{entanglement}
and theory of positive maps\index{map!positive}.

Finally, it should be mentioned that necessary conditions for separability%
\index{condition!for separability!necessary}
have been recently developed in infinite dimensions \cite{Giedke,Simon}.
In particular, the Peres criterion\index{criterion!Peres} was expressed
in terms of Wigner
representation and applied to Gaussian wave
packets\index{wave packet!Gaussian} \cite{Simon}.

\section{Entanglement and theory of positive maps}
\label{sec-ent-pos}

To describe the very fruitful connection between
entanglement\index{entanglement} and theory
of positive maps\index{map!positive} we will need mathematical notions like
positive operators\index{operator!positive},
positive maps\index{map!positive}, completely positive
maps\index{map!completely positive}. In the following section
we establish these notions. In next sections we will use them to
develop characterisation of the set of separable states\index{state!separable}.

\subsection{Positive and completely positive maps}
\label{sec-posCPmaps}
We start from the following notation.
By ${\cal A}_A$ and ${\cal A}_B$ we will denote the set of
operators\index{operator} acting
on ${\cal H}_A$ and ${\cal H}_B$ respectively.
Recall that the set ${\cal A}$ of
operators\index{operator} acting on some Hilbert space\index{space!Hilbert}
$\hcal$ constitute a Hilbert space\index{space!Hilbert}
itself (so-called Hilbert-Schmidt space\index{space!Hilbert-Schmidt})
with scalar product\index{product!scalar}
$\langle A,B\rangle=\trace A^\dagger B$.
One can consider an
operator\index{operator} orthonormal basis\index{basis!orthonormal}
in this space\index{space} given
by $\{|i\ra\la j|\}_{i,j=1}^{\dim \hcal}$ where $|i\ra$ is a
basis\index{basis}
in the space\index{space}
$\hcal$. Since we deal with finite dimension, $\acal$ is in fact
a space\index{space} of
matrices. Hence we will denote it sometimes by $M_d$ where $d$ is dimension of
$\hcal$.

The space\index{space} of the linear maps\index{map!linear}
from ${\cal A}_A$ to ${\cal A}_B$ is denoted by
${\cal L}({\cal A}_A,{\cal A}_B)$. We say that a map\index{map}
$\Lambda\in{\cal L}({\cal A}_A,{\cal A}_B)$
 is positive if it maps positive operators\index{operator!positive}
in ${\cal A}_A$ into the set of
positive operators\index{operator!positive} i.e.  if $A\geq0$ implies
$\Lambda(A)\geq0$. Finally we need the definition of completely positive (CP)
map\index{map!completely positive}. One says \cite{Kraus} that
a map\index{map} $\Lambda\in{\cal L}({\cal A}_A,{\cal
A}_B)$ is
completely positive if the induced map\index{map}
\begin{equation}
\Lambda_n=\Lambda\otimes \id_n:{\cal A}_A\otimes {\cal M}_n \rightarrow
{\cal A}_B\otimes {\cal M}_n
\end{equation}
is positive for all $n$; here $\id_n$ is the identity map\index{map!identity}
on the space\index{space} $\mcal_n$\footnote{Of course a
completely positive map\index{map!completely positive} is also a
positive one.}.
Thus the tensor product\index{product!tensor} of a
CP map\index{map!completely positive} and the
identity
maps positive
operators\index{operator!positive} into positive ones.  An example of
CP map\index{map!completely positive} is
$\varrho\rightarrow W\varrho W^\dagger$ where $W$ is an
arbitrary operator\index{operator}. As a matter of
fact, the general form of CP maps\index{map!completely positive} is
\be
\Lambda(\varrho)=\sum_i W_i\varrho W_i^\dagger\;.
\ee
CP maps\index{map!completely positive} that do not increase
trace ($\trace \Lambda(\varrho)\leq
\trace\varrho$) correspond to the most general physical operations%
\index{operation!physical}
allowed by
quantum mechanics\index{quantum!mechanics}  \cite{Kraus}.
If $\trace \Lambda(\varrho)=\trace \varrho$
for any $\varrho$ (we say the map is
trace preserving\index{map!trace preserving})
then the operation\index{operation}  can be performed with probability $1$,
otherwise with probability $p=\trace\Lambda(\varrho)$.

It is remarkable that there are positive
maps\index{map!positive}
that are not CP: an example is just
the transposition mentioned in the previous section.
Indeed, if $\varrho$ is positive, then so is $\varrho^T$, because
\be
\trace\varrho^T P= \trace \varrho P^T\geq0
\ee
and $P^T$ is still some projector\index{projector}. We used here the fact that
$\trace A^T=\trace A$. On the other hand $\id\otimes T$ is no longer positive.
One can easily check it, showing that $(\id\otimes T)P_+\equiv P_+^{T_B}$ 
is not a positive operator\index{operator!positive}.

A positive map\index{map!positive} is called {\it decomposable} \cite{Choi} if it can be
represented  in the form
\be
\Lambda=\Lambda^1_{CP}+\Lambda^2_{CP}\circ T\;,
\ee
where $\Lambda^i_{CP}$ are some CP maps\index{map!completely positive}.
For low dimensional systems\index{system!low-dimensional} ($\Lambda:M_2\rightarrow M_2$
or $\Lambda:M_3\rightarrow M_2$) the set of positive
maps\index{map!positive} can be easily
characterised. Namely  it has been shown \cite{Stormer,Woronowicz}
that {\it all} the positive maps\index{map!positive!decomposable} 
are decomposable in this case. If, instead, at
least one of the spaces\index{space} is ${\cal M}_n$ with $n\geq4$, there exist
non-decomposable
positive maps\index{map!positive!non-decomposable} \cite{Choi,Woronowicz}
(see example in Sect. \ref{examples}).
No full characterisation of positive
maps\index{map!positive} has been worked out so
far in this case.

\subsection{Characterisation of separable states {\it via}
positive maps}
\label{ent-pos3}
The fact that complete positivity is not equivalent to positivity is
crucial for the problem of entanglement\index{entanglement} we
discuss here. Indeed, trivially, the product
states\index{state!product} are mapped into positive
operators\index{operator!positive} by the tensor
product\index{product!tensor} of a positive
map\index{map!positive} and identity:
$(\Lambda\otimes \id)(\varrho\otimes\tilde\varrho)=
(\Lambda\varrho)\otimes\tilde\varrho\geq0$. Of course,	the same holds
for  separable
states\index{state!separable}. Then the main idea is  that this property of
the separable
states\index{state!separable} is essential i.e., roughly speaking, if
a state\index{state!entangled} $\varrho$ is
entangled, then there exists a
positive map\index{map!positive} $\Lambda$ such
that $(\Lambda\otimes \id)\varrho$ is {\it not}
positive.
This means that  one can seek the entangled
states\index{state!entangled} by means of the positive
maps\index{map!positive}.
Now  the point is that not all the positive maps\index{map!positive}
can help us to determine
whether a given state is
entangled\index{state!entangled}. In fact,  the completely positive
maps\index{map!completely positive} do not ``feel''
entanglement\index{entanglement}.
Thus the problem of characterisation of  the set of the separable
states\index{state!separable}
reduces to
the following: one should extract from that set of all
positive maps\index{map!completely positive}
some essential ones.
As we will see further, it is possible in some cases. Namely it appears that
for the $2\otimes2$ and $2\otimes3$
systems\index{system!$2\otimes 2$}\index{system!$2\otimes 3$} the transposition is the {\it only}
such map\index{map}. For higher dimensional
systems\index{system!higher-dimensional}, apart from transposition
also non-decomposable maps\index{map!positive!non-decomposable} will
be relevant.

Consider the lemma \cite{sep}  that will lead us to the basic theorem
relating entanglement\index{entanglement} and positive
maps\index{map!positive}
\begin{lemma}
A state $\varrho\in{\cal A}_A\otimes{\cal A}_B $ is
separable\index{state!separable} if and only if
\begin{equation}
\trace (A\varrho) \geq 0
\label{eqfunk}
\end{equation}
for any operator\index{operator} $A$ satisfying $Tr(A P\otimes Q) \geq 0$,
for all pure states\index{state!pure} P and Q acting on ${\cal H}_A$ 
and ${\cal H}_B$ respectively.
\label{hyper}
\end{lemma}

\textbf{Remark.} Note that
operator\index{operator} $A$ that is positive on
product states\index{state!product}
(i.e. satisfying $\trace A P\otimes Q\geq 0$) is automatically 
Hermitian \index{operator!Hermitian}.

The lemma is a reflection of the fact that in real Euclidean
space\index{space!real Euclidean},
a convex set and a point lying outside it can always be separated by
a hyper-plane\footnote{For infinite dimensions one must invoke Hahn-Banach
theorem\index{theorem!Hahn-Banach}, geometric form of
which is generalisation of this fact.}. Here, the
convex set is the set of separable
states\index{state!separable}, while the point is the
entangled state\index{state!entangled}. The hyper-plane is
determined by the
operator\index{operator} $A$.
The operator\index{operator} that is positive on
product states\index{state!product}  but is
not positive
has been called 
``entanglement witness''\index{entanglement!witness}\cite{Terhal-phd}, as
it indicates entanglement\index{entanglement} of 
some state\index{state} (first entanglement witness was provided 
in \cite{Werner}, see Sect. \ref{examples}). Now, to pass to
positive maps\index{map!positive}, we will use
isomorphism\index{isomorphism} between entanglement 
witnesses\index{entanglement!witness} and
positive non-CP maps\index{map!positive} \cite{Jamiolkowski}.
Note that if we have any linear operator\index{operator!linear}
$A\in\acal_A\otimes\acal_B$, we can
define a map\index{map} $\Lambda\in\lcal(\acal_A,\acal_B)$ by
\be
\la k|\,\Lambda(|i\ra\la j|)\,|l\ra= \la i|\otimes\la k|\,A\,|j\ra\otimes| l\ra
\ee
which can be rephrased as follows
\be
{1\over d}A=(\id\otimes \Lambda)\,P_+^d\;
\label{witness-map}
\ee
with $d=\dim\hcal_A$.
Conversely, given a map\index{map}, the above formula allows to obtain a
corresponding
operator\index{operator}. It turns out, that this formula gives also
one-to-one correspondence
between entanglement witnesses\index{entanglement!witness}
and  positive non-CP maps\index{map!positive}
\cite{Jamiolkowski}.  Applying this fact one can  prove \cite{sep}
 the following theorem

\begin{theorem}
Let $\varrho$ act on Hilbert
space\index{space!Hilbert} ${\cal H}_A \otimes {\cal H}_B$. Then
$\varrho$ is separable if and only if for any positive
map\index{map!positive}
$\Lambda:{\cal A}_B\rightarrow {\cal A}_A$ the operator\index{operator}
$(\id\otimes\Lambda)\varrho$ is positive.
\label{dod}
\end{theorem}
As we mentioned, the relevant positive
maps\index{map!positive} are here the  ones that are not
completely positive. Indeed, for
CP map\index{map!completely positive} $\Lambda$ we have $(\id\otimes
\Lambda)\varrho\geq0$ for any state\index{state}  $\varrho$, hence CP
maps\index{map!completely positive} are of no use
here.
The above theorem presents, to authors knowledge, the first application of the
theory of positive maps\index{map!positive}  in physics. So far, only completely positive ones
were  of interest for physicists. As we will see the theorem proved
fruitful both for mathematics (theory of
positive maps\index{map!positive}) and for physics
(theory of entanglement\index{entanglement}).

\subsection{Operational characterisation of entanglement in  low dimensions
($2\otimes2$ and $2\otimes 3$ systems)}
The first conclusion from the theorem is operational characterisation of the
separable  states\index{state!separable} in low
dimensions ($2\otimes 2$ and $2\otimes 3$). It
follows from the mentioned fact that
positive maps\index{map!positive} in low dimensions
are decomposable. Then the condition $(\id\otimes \Lambda)\varrho\geq0$ reads
as $(\id\otimes\Lambda_1^{CP})\varrho+(\id\otimes\Lambda^{CP}_2)
\varrho^{T_B}$. Now,
since $\varrho$ is positive and $\Lambda_1^{CP}$ is CP, the
first term is always positive. If $\varrho^{T_B}$ is positive, then also
the second term is positive, hence their sum is a positive
operator\index{operator!positive}. Thus
to check whether for all positive
maps\index{map!positive} we have $(\id\otimes \Lambda) \varrho\geq0$
it suffices to check only transposition. One obtains \cite{sep} 
(see \cite{Osaki} in this context)
\begin{theorem}
A state\index{state} $\varrho$ of  $2\otimes 2$ or $2\otimes 3$
system\index{system!$2\otimes 2$}\index{system!$2\otimes 3$} is separable
if and only if its partial transposition is a
positive operator\index{operator!positive}.
\label{2x2}
\end{theorem}

\textbf{Remark.} Equivalently one can use the partial transposition with
respect to the first space\index{space}.

The above theorem is an important result, as it allows	to determine
unambiguously whether a given quantum state\index{quantum!state}
of $2\otimes2$ ($2\otimes 3$)
system\index{system!$2\otimes 2$}\index{system!$2\otimes 3$} can be written as mixture\index{mixture} of
product states\index{state!product} or not.
The necessary and sufficient condition for separability%
\index{condition!for separability!necessary and sufficient}
is here surprisingly
simple, hence it found many applications.
In particular, it was applied in the context of  broadcasting entanglement%
\index{broadcasting!entanglement}
\cite{broad}, quantum information\index{information!quantum}
\index{quantum!information} flow in
quantum copying networks
\cite{net},
disentangling machines\index{disentangling machine} \cite{disent},
imperfect two-qubit\index{qubit} gate\index{gate!two-qubit} \cite{Poyatos},
analysis of volume of the set of\index{volume!of the set of entangled states}
entangled states\index{state!entangled}
 \cite{volume,Karol}, decomposition of
separable states\index{decomposition!of separable state}\index{state!separable} into minimal
ensembles or pseudo-ensembles \cite{Sanpera},
entanglement\index{entanglement!splitting}
 splitting \cite{Dagmar}, analysis of entanglement measures%
 \index{entanglement!measure}
\cite{Wooters,Tarrach,Eisert}.

In Sect. \ref{sec-two-qubits} we describe the first application \cite{pur}: by
use of the theorem we show that any entangled two-qubit\index{qubit}
system can\index{system!two-qubit}
be distilled, hence is useful for quantum communication%
\index{communication!quantum}\index{quantum!communication}.

\subsection{Higher dimensions - entangled states with positive partial
transposition}
\label{PPT-ent}
Since the St\o{}rmer-Woronowicz characterisation
of positive maps\index{map!positive} applies only to low
dimensions, it follows that for higher dimensions partial transposition will
not constitute necessary and sufficient condition for separability%
\index{condition!for separability!necessary and sufficient}. Thus there
exist states that are entangled\index{state!entangled}, but
are PPT\index{state!product} (see Fig. \ref{fig-ent})
\begin{figure}
\ \vskip1.5cm
\parbox{200pt}{
\begin{picture}(200,150)
\put(28,175){(a)}
\put(150,100){\oval(240,120)}    
\put(150,40){\line(0,1){120}}    
\put(80,170){\bf PPT}
\put(200,170){\bf NPT}
\put(61,100){\parbox{2.6cm}{\large \bf separable \\ \centerline{states\ 
\hskip1mm  \ \ \ }}}
\put(181,100){\parbox{2.6cm}{\large \bf entangled\\ \centerline{states \ \ \ }}}
\end{picture}}

\vskip1cm

\parbox{200pt}{
\begin{picture}(200,150)
\put(28,175){(b)}
\put(150,100){\oval(240,120)}    
\put(150,40){\line(0,1){85}}    
\put(150,150){\line(0,1){10}}    
\put(92,96){\oval(80,50)}
\put(80,170){\bf PPT}
\put(200,170){\bf NPT}
\put(61,95){\ \parbox{2.4cm}{\large \bf separable \\ \centerline{states \ \ \ }}}
\put(100,135){\parbox{5cm}{\large \bf entangled states }}
\end{picture}}
\caption[Structure of entanglement of mixed states]{Structure of
entanglement of mixed states for $2\otimes 2$ and $2\otimes3$ system (a) and
for higher dimensions (b)}
\label{fig-ent}
\end{figure}
First explicit examples of a  entangled but
PPT state\index{state!PPT} were provided in
\cite{transp}. Later on it appeared, that the mathematical literature
concerning non-decomposable maps\index{map!positive!non-decomposable} contains
examples of matrices that can be treated as prototypes of PPT entangled 
states\index{state!PPT entangled}  \cite{Osaki,Stormer2}.

We will now describe the way of obtaining them presented in
\cite{transp}, as it proved to be a fruitful direction in searching for
PPT entangled states\index{state!PPT entangled}. Chapter \ref{chap-dist} will provide motivation
for undertaking the very tedious task of the search -- the
states\index{state}
will represent a curious type of entanglement%
\index{entanglement}  -- bound
entanglement\index{entanglement!bound}.

To find desired examples we must take a PPT state\index{state!PPT} and somehow show that it is
entangled. Of course, we cannot use the strongest so far tool, i.e. PPT
criterion\index{criterion!PPT}, just because the state is to be PPT. So we
must derive a
criterion that
would be {\it stronger} in some cases. It appears that the very
range\index{range}%
\footnote{The range\index{range} of an
operator\index{operator} $A$ acting on the Hilbert 
space\index{space!Hilbert} $\hcal$ is given by 
$R(A)=\{A(\psi):\psi\in\hcal\}$.
If $A$ is
Hermitian operator\index{operator!Hermitian} then the range is equivalent 
to the support, i.e., the space\index{space} spanned by its
eigenvectors\index{eigenvectors} with nonzero eigenvalues\index{eigenvalues}.
}
of the state\index{state} can say us much about
its entanglement\index{entanglement} in some cases. This is
contained in the following theorem, derived in \cite{transp} on the basis of
analogous condition for positive maps\index{map!positive!non-decomposable}
considered in  \cite{Woronowicz}.

\begin{theorem}{\it (range criterion)}

\noindent
If a state\index{state} $\varrho$ acting on the space\index{space} $\hcal_{AB}$
is separable, then there exists a family of product
vectors\index{vector!product}
$\psi_i\otimes \phi_i$ such that
\bei
\item[a)] they span the range\index{range} of $\varrho$
\item[b)] the
vectors\index{vector}  $\{\psi_i\otimes\phi_i^*\}_{i=1}^{k}$ span  the
range\index{range}
of $\varrho^{T_B}$ (where $*$ denotes complex conjugation in the  basis%
\index{basis} in which partial transposition was performed).
\eei
In particular, any of
the vectors\index{vector} $\psi_i\otimes \phi_i^*$ belongs to the
range\index{range} of $\varrho$.
\label{thPawel}
\end{theorem}

Now, in   \cite{transp} there were presented two examples of PPT states%
\index{state!PPT}
violating the above criterion. We will present the
example for $2\otimes 4$ case%
\footnote{It bases on  an example concerning positive 
maps\index{map!positive} \cite{Woronowicz}.}.
The matrix is written in the standard product
basis\index{basis!product} $\{|ij\ra\}$
\begin{eqnarray}
\varrho_b={1 \over 7b + 1}
\left[ \begin{array}{cccccccc}
	   b&0&0&0&0&b&0&0   \\
	   0&b&0&0&0&0&b&0     \\
	   0&0&b&0&0&0&0&b     \\
	   0&0&0&b&0&0&0&0     \\
	   0&0&0&0&{1+b \over 2}&0&0&{\sqrt{1-b^2} \over 2} \\
	   b&0&0&0&0&b&0&0     \\
	   0&b&0&0&0&0&b&0     \\
	   0&0&b&0&{\sqrt{1-b^2} \over 2}&0&0&{1+b \over 2}\\
       \end{array}
      \right ]\;,
\end{eqnarray}
where $0<b<1$. Now performing PT, as shown by formula (\ref{PT-wykonanie}) we
can check  that $\varrho_b^{T_B}$ remains
positive operator\index{operator!positive}. By a tedious
calculations, one can check, that {\it none} of the product
vectors\index{vector!product} belonging
to the range\index{range} of $\varrho_b$, if partially
conjugated (as stated in theorem),
belong to the range\index{range} of $\varrho_b^{T_B}$. Thus
the condition stated in
theorem is drastically violated, hence the
state is entangled\index{state!entangled}. As we will see
further, the entanglement\index{entanglement} is masked so subtly,
that it cannot be distilled
at all!

\subsubsection{Range criterion and positive non-decomposable maps}
\label{sec-range}


The separability criterion\index{criterion!separability} given by the above theorem has been  fruitfully
applied in search for PPT entangled states\index{state!PPT entangled}
\cite{UPB1,UPB2,Peres_Bruss}.
The Theorem \ref{thPawel} was applied in \cite{Lew_San} where 
a technique of subtraction of product vectors\index{vector!product}
from the range\index{range} of 
state\index{state} was used to get best separable approximation (BSA)
\index{best separable approximation (BSA)} of the state \index{state}.
As a tool, the authors considered subspaces\index{subspace}
containing {\it no}
product vectors\index{vector!product}. Note that 
 the (normalised) projector\index{projector}
onto such a subspace\index{subspace} must be entangled, as the condition a)
from the theorem is not satisfied. This approach  was successfully 
applied in Ref. \cite{UPB1}  
(see also \cite{UPB2,Terhal-phd}) and, in connection
with seemingly completely different concept of {\it unextendible product
bases}\index{basis!product!unextendible (UPB)}, produced an elegant, and so
far the most transparent way of
construction of PPT entangled states\index{state!PPT entangled}.

To describe the construction\footnote{The construction applies to
multipartite case \cite{UPB1}, in the present review we consider only
bipartite systems.}, one needs  the following definition \cite{UPB1}

\begin{definition}
A set of product orthogonal
vectors\index{vectors!orthogonal}\index{vector!product} in $\hcal_{AB}$
\bei
\item[a)] that has less elements than the dimension of the
space\index{space}
\item[b)] such that there does not exist
any product vector orthogonal\index{vector!product}%
\index{vectors!orthogonal} to all of them
\eei
is called {\it unextendible
product basis}\index{basis!product!unextendible (UPB)}.
\end{definition}

Here we recall an example of such basis\index{basis} in
$3\otimes 3$ system\index{system!$3\otimes 3$}:
\beq
|v_0\ra={1\over \sqrt2}|0\ra|0-1\ra,\quad
|v_2\ra={1\over \sqrt2}|2\ra|1-2\ra,\nonumber\\
|v_1\ra={1\over \sqrt2}|0-1\ra|2\ra,\quad
|v_3\ra={1\over \sqrt2}|1-2\ra|0\ra\\
|v_4\ra={1\over 3}|0+1+2\ra|0+1+2\ra\;.\nonumber
\eeq
Of course, the above five vectors are
orthogonal\index{vectors!orthogonal} to each other. However, any subset
of three vectors\index{vectors} on either side spans the full three-dimensional
space\index{space}.
This prevents from existence of a sixth product
vector\index{vector!product} that would be
orthogonal to all five of them. How to connect this with the problem 
we deal in
this section? The answer is: via  the subspace\index{space} 
complementary to the one spanned by these
vectors\index{vector}. Indeed, suppose that $\{w_i=\phi_i\otimes
\psi_i\}^k_{i=1}$ is UPB\index{basis!product!unextendible (UPB)}. 
For $d\otimes d$
system\index{system!$d\otimes d$}. Consider projector\index{projector}
$P=\sum_{i=1}^k|w_i\ra\la w_i|$ onto the subspace\index{subspace} $\hcal$ spanned by
the vectors\index{vector}
$w_i$ ($\dim\hcal=k$)
Now, consider the state\index{state} uniformly distributed on its orthogonal
complement $\hcal_\perp$ ($\dim\hcal_\perp=d^2-k$)
\be
\varrho={1\over d^2-k} (\id-P)\;.
\label{UPB-bound}
\ee
The range\index{range} of the state\index{state} ($\hcal_\perp$)
contains no product vectors\index{vector!product}:
otherwise one would be able to extend the product basis\index{basis!product}
$\{w_i\}$. Then by
Theorem \ref{thPawel} the state\index{state} must be entangled. Let us now calculate
$\varrho^{T_B}$. Since $w_i=\phi_i\otimes \psi_i$ then
$(|w_i\ra\la w_i|)^{T_B}=|\tilde w_i\ra\la\tilde w_i|$ where
$\tilde w_i=\phi_i\otimes \psi_i^*$. The vectors\index{vector} $\tilde w_i$ are
orthogonal to each other so that the operator\index{operator}
$P^{T_B}=\sum_i|\tilde w_i\ra\la \tilde w_i|$ is a
projector\index{projector}. Consequently,
the operator\index{operator} $(\id-P)^{T_B}=\id -P^{T_B}$ is also
projector\index{projector}, hence it
is positive. We conclude that $\varrho$ is PPT.

A different way of obtaining examples of PPT
entangled states\index{state!PPT entangled}
can be inferred from the papers
devoted to search for non-decomposable positive
maps\index{map!positive!non-decomposable} in mathematical
literature \cite{Osaki,Stormer2} (see Sect. \ref{sec-posCPmaps}). 
A way to find non-decomposable map
\index{map!positive!non-decomposable}
is the following. One constructs some
map\index{map} $\Lambda$ and proves somehow that
it is positive.
Thus one must guess some (possibly unnormalised)
state\index{state} $\varrho$ that
is PPT. Now, if $(\id\otimes \Lambda)\varrho$ is not positive,
then $\Lambda$ cannot be decomposable, as shown in the discussion
preceding Theorem \ref{2x2}. At the same time, the
state\index{state} must be entangled.
In Sect. \ref{examples} we present an example of PPT
entangled state\index{state!PPT entangled}
(based on \cite{Stormer}) found in this way. Thanks to its symmetric form
the state\index{state} allowed to reveal the first quantum effect produced by bound
entanglement\index{entanglement!bound} (see Sect. \ref{aktywacja}).

Thus a possible direction of exploring the ``PPT region'' of entanglement%
\index{entanglement}
is to develop the description of non-de\-com\-po\-sa\-ble
maps\index{map!positive!non-decomposable}.
However, it appears
that there can be also a ``back-reaction'': exploration of PPT region, allowed
to obtain new results  on
non-decomposable maps\index{map!positive!non-decomposable}.
It turns out that  just the UPB\index{basis!product!unextendible (UPB)} 
method  described before allows for easy  construction of new
non-decomposable
maps\index{map!positive!non-decomposable} \cite{Terhal}. The
interested reader we direct to
the original article as well as \cite{Terhal-phd}. We only note that
to find a
non-decomposable map\index{map!positive!non-decomposable}, one
needs only to construct some UPB\index{basis!product!unextendible (UPB)}.
Then the procedure is automatic, as the one described above. 
To our knowledge, this is
the first {\it systematic} way of finding
non-decomposable maps\index{map!positive!non-decomposable}.

\section{Examples}

\label{examples}
We present here a couple of examples, illustrating the results contained
in previous sections. In particular we introduce two families of
states\index{state} that
play important role in the problem of distillation of entanglement%
\index{entanglement!distillation}.

\subsection{Reduction criterion for separability}
As mentioned in Sect. \ref{ent-pos3}, if $\Lambda$ is a
positive map\index{map!positive} then
for separable states\index{state!separable} we have
\be
(\id\otimes \Lambda) \varrho\geq0\;.
\label{nec-sep}
\ee
If the map\index{map} is not CP, then this condition is not
trivial, i.e. for some states\index{state}
$(\id\otimes \Lambda) \varrho$ is not positive. Consider the
map\index{map} given by
$\Lambda(A)=(\trace A) \id -A$. The eigenvalues\index{eigenvectors} of
the resulting operator\index{operator}
$\Lambda(A)$ are given by $\lambda_i=\trace A -a_i$ where $a_i$ are
eigenvalues\index{eigenvalues} of $A$. If $A$ is positive, then $a_i\geq0$.
Now, since $\trace A=\sum_i a_i$ then also $\lambda_i$ are nonnegative.
Thus the map\index{map} is positive. Now,
the formula (\ref{nec-sep}) and
the dual one
$(\Lambda\otimes \id)\varrho\geq0$  applied to this particular
map\index{map}
implies that separable states\index{state!separable} must satisfy the following inequalities
\be
\id\otimes \varrho_B-\varrho\geq0,\quad \varrho_A\otimes \id -\varrho\geq0\;.
\ee
The two conditions taken jointly are called reduction
criterion\index{criterion!reduction}
\cite{xor,cerf}. One can check that it implies the entropic
inequalities\index{inequalities!entropic}
(hence it is better in ``detecting'' entanglement\index{entanglement}). From
the reduction
criterion\index{criterion!reduction} it follows that
states\index{state} $\varrho$ of
$d\otimes d$ system\index{system!$d\otimes d$}
with $\fcal(\varrho)>1/d$ must be entangled (this was originally argued in
\cite{huge}). Indeed, from the above inequalities
it follows that for separable state\index{state!separable} $\sigma$ and a maximally
entangled state\index{state!entangled!maximally}
$\psi_{me}$  one has
$\la\psi_{me}|\sigma_A\otimes \id-\sigma|\psi_{me}\ra\geq0$. Since the reduced
density matrix\index{matrix!density!reduced} $\varrho^{\psi_{me}}_A$ of the
state $\psi_{me}$ is
proportional to identity we obtain $\la\psi_{me}|\sigma_A\otimes
\id|\psi_{me}\ra=\trace(\varrho^{\psi_{me}}_A\sigma_A)=1/d$. Hence we get
$\fcal\leq 1/d$. We conclude that the latter condition is separability
criterion\index{criterion!separability}.

Let us finally note \cite{xor}, that for $2\otimes 2$ and $2\otimes 3$
systems\index{system!$2\otimes 2$}\index{system!$2\otimes 3$}
the reduction criterion\index{criterion!reduction} is equivalent to PPT
criterion\index{criterion!PPT}, hence is equivalent to separability.

\subsection{Strong separability criteria from entanglement
witness}
Consider unitary {\it flip}  operator\index{operator!flip} $V$
on $d\otimes d$ system\index{system!$d\otimes d$} defined by
$V\psi\otimes \phi=\phi\otimes \psi$.
Note that it can be written as $V=P_{S}-P_{A}$ where $P_{S}$ ($P_A$) is
projector\index{projector} onto symmetric (antisymmetric)
subspace\index{subspace!symmetric}\index{subspace!antisymmetric}
of the total space\index{space}.
Hence $V$ is dichotomic observable\index{observable!dichotomic} (with
eigenvalues\index{eigenvalues}
$\pm1$). One can
check that $\trace VA\otimes B=\trace AB$ for any
operators\index{operator} $A,B$. Then
$V$ is an entanglement witness\index{entanglement!witness}, so that
$\trace\varrho V\geq 0$ is
separability criterion\index{criterion!separability} \cite{Werner}. Now,
let us find the corresponding
positive map\index{map!positive} via the
formula (\ref{witness-map}). One easily gets
that it is transposition (up to an irrelevant factor). Remarkably,
in this way, given  entanglement witness\index{entanglement!witness}, one
can find the
corresponding map\index{map}, to obtain {\it much stronger} criterion given by formula
(\ref{nec-sep}).

\subsection{Werner states}
In \cite{Werner} Werner considered states\index{state} that do not change 
if subjected to
the same unitary transformation\index{transformation!unitary} to both
subsystems\index{subsystem}:
\be
\varrho=U\otimes U\,\varrho\, U^\dag\otimes U^\dag \quad \mbox{for any
unitary } U\;.
\ee
He showed that such states (called Werner states)\index{state!Werner} must be
of the following form
\be
\varrho_W(d)={1\over d^2-\beta d} (\id+\beta V), \quad -1\leq\beta\leq1\;,
\label{Werner-state}
\ee
where  $V$ is flip operator\index{operator!flip} defined above.
Other form for $\varrho_W$ is \cite{LewNPT}
\be
\varrho_W(d)=p{P_{A}\over N_A}+(1-p){P_S\over N_S}, \quad 0\leq p\leq1\;,
\ee
where $N_A=(d^2-d)/2$ ($N_S=(d^2+d)/2$) is the dimension of the antisymmetric
(symmetric) subspace\index{subspace!symmetric}\index{subspace!symmetric}.
It was shown \cite{Werner} that $\varrho_W$ is entangled if and only if
$\trace V\varrho_W<0$. Equivalent conditions are: $\beta<-1/d$, $p>0$ or
$\varrho$ is NPT. Thus $\varrho_W$ is separable if and only if
it is PPT. For $d=2$ (two-qubit\index{qubit} case)
the state\index{state} can be written as (see \cite{Popescu94} in this context)
\be
\varrho_W(2)=p|\psi_-\ra\la\psi_-|+(1-p) {\id\over 4}, \quad -{1\over
3}\leq p\leq1\;.
\label{Wer-Pop}
\ee

Note that any state\index{state} $\varrho$ if subjected
to random transformation\index{transformation!random}
of the form $U\otimes U$ (call such operation\index{operation!twirling}
 $U\otimes U$ {\it twirling})\index{twirling!$U\otimes U$}
becomes Werner state\index{state!Werner}
\be
\int {\rm d\,} U \,U\otimes U\,\varrho\, U^\dagger\otimes U^\dagger=\varrho_W\;.
\ee
Moreover  $\trace\varrho V=\trace\varrho_W V$  (i.e. $\trace \varrho V$ is
invariant of $U\otimes U$ twirling\index{twirling!$U\otimes U$})).

\subsection{Isotropic state}
If we apply   local unitary transformation\index{transformation!unitary!local} to the
state\index{state} (\ref{Wer-Pop}), changing
$\psi_-$ into $\psi_+$ we can generalise its form  to higher dimension
as follows \cite{xor}
\be
\varrho(p,d)=
p P_+ +  (1-p){\id\over d^2},\quad \ \rm{with} \quad
-{1\over d^2-1} \leq p \leq 1\;.
\label{isotropic}
\ee
The state\index{state!isotropic} will be
called isotropic \cite{Rains}%
\footnote{In \cite{xor} it was called noisy singlet.}.
For $p>0$ it is interpreted as  mixture\index{mixture} of
maximally entangled state\index{state!entangled!maximally}
$P_+$ with a completely chaotic noise represented by $\id/d$.
It was shown that it is the only state\index{state}
invariant under $U\otimes U^*$
transformations\index{transformation!$U\otimes U^*$}\footnote{
The star denotes complex
conjugation.}. If we use singlet fraction\index{fraction!singlet}
$F=\trace\varrho P_+$ as a parameter,
we obtain
\be
\varrho(F,d)={d^2\over d^2-1}\left((1-F){\id\over d^2}+
(F-{1\over d^2})P_+\right),\quad 0\leq F\leq 1\;.
\ee
The two parameters are related via $p=(d^2F-1)/(d^2-1)$. The
state\index{state} is
entangled if and only if $F>1/d$, or equivalently if it is NPT.
Similarly  as for Werner states\index{state!Werner}, a state
subjected to $U\otimes U^*$ twirling\index{twirling!$U\otimes U^*$}
\index{twirling}
(random $U\otimes U^*$ operations\index{operations!random}%
\index{operation!$U\otimes U^*$})
becomes isotropic\index{state!isotropic}, and 
the parameter $F(\varrho)$ is invariant
under this operation.

\subsection{A two-qubit state}
Consider the following two-qubit\index{qubit} state\index{state}
\be
\varrho=p|\psi_-\ra\la\psi_-| +(1-p)|00\ra\la00|\;.
\label{przyklad}
\ee
By formula (\ref{mbell}) we obtain that for $p\leq 1/\sqrt2$ CHSH-Bell
inequalities\index{inequalities!CHSH-Bell} are satisfied. A
little bit stronger is the criterion
involving fully entangled fraction\index{fraction!fully entangled}, we
have $\fcal\leq 1/2$ for
$p\leq 1/2$.
Entropic inequalities\index{inequalities!entropic}, apart from the one
involving $S_0$,  are
equivalent to each other for this state\index{state}
and give again $p\leq 1/2$. Thus they reveal
entanglement\index{entanglement} for $p>1/2$.
Applying the partial transposition one can convince, that the
state\index{state} is
entangled for all $p>0$ (for $p=0$ it is manifestly separable).

\subsection{Entangled PPT state via non-decomposable positive map}
Consider the following state\index{state} (constructed on the basis
of St\o rmer
matrices \cite{Stormer}) of
$3\otimes 3$ system\index{system!$3\otimes 3$}
\be
\sigma_{\alpha}=\frac{2}{7}|\psi_{+}\rangle \langle \psi_{+}|
+\frac{\alpha}{7} \sigma_{+}+
\frac{5-\alpha}{7} \sigma_{-}\quad 2\leq\alpha\leq 5\;.
\label{eq-Stormer}
\end{equation}
with
\beq
\sigma_{+}={1 \over 3}(|0\rangle|1\rangle \langle 0| \langle 1|
+ |1\rangle|2\rangle \langle 1| \langle 2|+
|2\rangle|0\rangle \langle 2| \langle 0|)\;,
\label{mix+}\nonumber\\
\sigma_{-}={1 \over 3}(|1\rangle|0\rangle \langle 1| \langle 0|
+ |2\rangle|1\rangle \langle 2| \langle 1|+
|0\rangle|2\rangle \langle 0| \langle 2|)\;.
\label{mix-}
\eeq
Using the formulas (\ref{PT-wykonanie}) one easily finds that for
$\alpha\leq4$ the state is PPT\index{state!PPT}. Consider now the following map\index{map}
\cite{Choi}:
\be
\Lambda\left(\left[
\ba{ccc}
a_{11} &a_{12} & a_{13}\\
a_{21} &a_{22} & a_{23}\\
a_{31} &a_{32} & a_{33}\\
\ea
\right]\right)=
\left[
\ba{ccc}
a_{11} &-a_{12} & -a_{13}\\
-a_{21} &a_{22} & -a_{23}\\
-a_{31} &-a_{32} & a_{33}\\
\ea
\right]+
\left[
\ba{ccc}
a_{33} &0 & 0\\
0 &a_{11} & 0\\
0 &0 & a_{22}\\
\ea
\right]\;.
\ee
This map\index{map} was shown to be positive \cite{Choi}. Now, one can
calculate the
operator\index{operator} $(\id\otimes \Lambda)\varrho$ and find
that one of its
eigenvalues\index{eigenvalues} is
negative for $\alpha>3$ (explicitly $\lambda_-=(3-\alpha)/2$). This implies
that
\bei
\item the state is entangled\index{state!entangled} (for separable state
\index{state!separable} we would have $(\id\otimes
\Lambda)\varrho\geq0$)
\item the map is
non-decomposable\index{map!positive!non-decomposable}
(for decomposable map\index{map!positive!decomposable} and
PPT state\index{state!PPT} we also
would have $(\id\otimes \Lambda)\varrho\geq0$).
\eei
For $2\leq\alpha\leq3$ it is separable, as it can be written as a
mixture\index{mixture}
of other separable states\index{state!separable}
$\sigma_{\alpha}=\frac{6}{7}\varrho_{1}
+\frac{\alpha-2}{7} \sigma_{+}+ \frac{3-\alpha}{7} \sigma_{-}$ with
$\varrho_1=(|\psi_{+}\rangle \langle \psi_{+}|+
\sigma_{+}+\sigma_{-})/3$. The latter state\index{state} can be written as an integral
over product states\index{state!product}:
\[
\varrho_{1}=\frac{1}{8}\int_{0}^{2\pi}
| \psi(\theta) \rangle \langle \psi(\theta)
| \otimes | \psi(-\theta) \rangle \langle \psi(-\theta) |
\frac{{\rm d\,}\theta}{2\pi}\;,
\]
with 
$| \psi(\theta) \rangle=
\frac{1}{\sqrt{3}}(|0\ra+e^{i\theta}|1\ra+e^{-2i \theta}|2\ra)$
(there exists also finite decomposition%
\index{decomposition!of separable state} exploiting phases of roots
of unity \cite{doktor}).

\section{Volumes of entangled and separable states}
\label{vol}
The question of volume of the set of separable or entangled
states\index{volume!of the set of entangled states}%
\index{volume!of the set of separable states}%
\index{state!separable} \index{state!entangled} in
the set of all states\index{state} raised in \cite{volume} is important for several reasons. First
one could be interested in the following basic question: Is the
world more\index{world!classical}\index{world!quantum}
classical or more quantum? Second, the size of the
volume\index{volume} would reflect the
fact important for numerical analysis of entanglement\index{entanglement}, to
what extent the
separable or entangled
states\index{state!entangled}\index{state!separable} are typical. Later it appeared that
the considerations on volume of separable
states\index{volume!of the set of separable states}%
\index{state!separable} lead of important results
concerning the question of relevance of entanglement\index{entanglement}
in quantum computing\index{quantum!computing}
\cite{nmr}.

We will mainly consider qualitative question: is the volume of
separable ($V_s$), entangled ($V_{e}$) or PPT entangled ($V_{pe}$)%
\index{volume!of the set of PPT states}%
\index{volume!of the set of separable states}%
\index{volume!of the set entangled states}
states\index{state!PPT entangled}\index{state!entangled}%
\index{state!separable} nonzero?
All these problems can be solved by the same method \cite{volume}:
one picks a suitable state\index{state}  from either of the sets,
and tries to show that some
(perhaps small) ball round the state\index{state}   is still contained
in the set.

For separable states\index{state!separable}  one takes the ball
round the maximally mixed state\index{state!mixed!maximally}:
one needs a number $p_0$ such that for any
state\index{state}  $\tilde\varrho$ the state\index{state}
\be
\varrho=p \id/N+(1-p) \tilde\varrho
\label{mieszanka}
\ee
is separable for all $p\leq p_0$ (here $N$ is the dimension of the total
system\index{system}). In \cite{volume} it was shown that
in the general case of
multipartite systems\index{system!multipartite} of any
finite dimension, such $p_0$  exists.
Note that, in fact, we obtained {\it sufficient} condition for separability%
\index{condition!for separability!sufficient}:
if the eigenvalues\index{eigenvalues} of a given state\index{state}
 do not differ too much
from the uniform spectrum\index{spectrum} of the maximally
mixed state\index{state!mixed!maximally}%
, then the state must be separable\index{state!separable}.
One would like to have some concrete values of $p_0$ that constitute
the condition (the larger $p_0$, the stronger the condition).

Consider, for example,
$2\otimes 2$ system\index{system!$2\otimes 2$}. Here one can provide the largest
possible $p_0$, as there exists the necessary and sufficient condition
for separability%
\index{condition!for separability!necessary and sufficient}
(PPT criterion)\index{criterion!PPT}. Consider
eigenvalues\index{eigenvalues} $\lambda_i$ of the
partial transposition of state\index{state}
(\ref{mieszanka}). They are of the form $\lambda_i=(1-p)/N+ p\tilde\lambda_i$
where $\tilde \lambda_i$ are eigenvalues\index{eigenvalues} of $\tilde\varrho^{T_B}$
(in our case $N=4$).
One easily can see (basing on the Schmidt decomposition%
\index{decomposition!Schmidt})  that
 partial transposition of a pure state\index{state!pure} cannot have
 eigenvalues\index{eigenvalues} smaller than
 $-1/2$. Hence the same is true for mixed states\index{state!mixed}. In
 conclusion
we obtain that  if $(1-p)/N-p/2\geq0$ then the
eigenvalues\index{eigenvalues} $\lambda_i$ are
nonnegative for arbitrary  $\tilde\varrho$. Thus
for $2\otimes2$ system\index{system!$2\otimes 2$}
one can take $p_0=1/3$ to obtain  sufficient condition for separability%
\index{condition!for separability!sufficient}.
Concrete values of $p_0$ for the case of
n-partite systems\index{system!n-partite} each
of dimension $d$ were obtained in \cite{Tarrach}:
\be
p_0={1\over (1+{2\over d})^{n-1}}.
\ee
These considerations proved to be crucial for analysis of the experimental
implementation
of quantum algorithm\index{quantum!algorithm} in high
temperature systems\index{system!high temperature}  via
nuclear magnetic resonance (NMR)\index{nuclear magnetic resonance} methods.
This is because a generic state used in this approach is the maximally mixed
one with a small admixture  of some pure entangled
state\index{state!pure entangled}.
In \cite{nmr} the sufficient  conditions%
\index{condition!for separability!sufficient} of the above sort were further
developed and  it was concluded,
that in all the NMR quantum computing\index{quantum!computing}
\index{nuclear magnetic resonance!quantum computing}
experiments performed so far
the admixture of the pure  state\index{state!pure} was too small.  Thus
the total state\index{state} used in these experiments was separable: it
satisfied condition sufficient for separability%
\index{condition!for separability!sufficient}. This raised
an interesting discussion
to what extent entanglement\index{entanglement} is necessary for
quantum computing\index{quantum!computing}
\cite{LindPop,Schack} (see also \cite{Knill} in this context).
Even though there is still no general answer, it was shown \cite{LindPop}
that Shor algorithm \cite{Shor_alg} requires
entanglement\index{entanglement}.

Let us now turn back  to the question of
volumes\index{volume} of $V_{e}$ and
$V_{pe}$. If one takes $\psi_+$ of a $d\otimes d$
system\index{system!$d\otimes d$} for simplicity,
it is easy to see that a not very large admixture of any
state\index{state} will keep
$F>1/d$. Thus any state\index{state} belonging to the neighbourhood
must be entangled.
Showing that the volume of PPT entangled states%
\index{volume!of the set of PPT entangled states}%
\index{state!PPT entangled} is
nonzero is a bit more
involved \cite{volume}.

\begin{figure}
\includegraphics[angle=270,width=.5\textwidth]{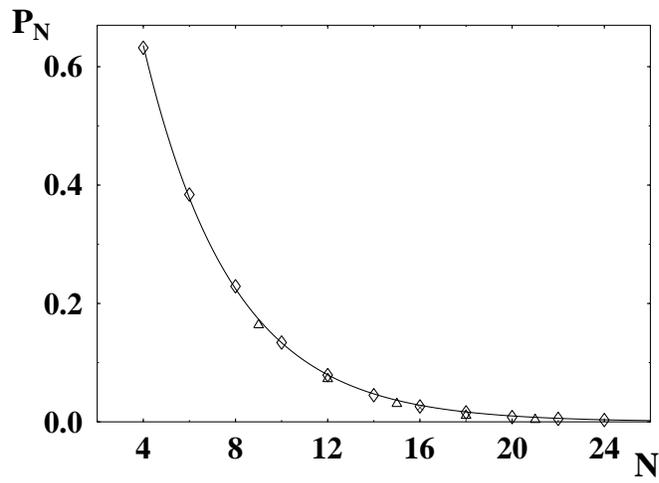}
\caption[Relative volume of the set of PPT states]{The ratio
$P_N=V_{PPT}/V$ of the volume o PPT states to the volume of
the set of all states versus dimension $N$ of the total system\index{system}.
Different symbols distinguish different sizes of
one subsystem\index{subsystem} ($k=2$
($\diamond$), $k=3$ ($\scriptstyle\bigtriangleup$)). (This figure 
is reproduced from Phys. Rev. A {\bf 58}, 883 (1998) by 
permission of authors.)} 
\label{fig-vol}
\end{figure}

In conclusion, all the three types of states\index{state} are not atypical
in the set of all states\index{state} of a
given system\index{system}. However, it appears that the
 ratio of the volume set of PPT states%
 \index{volume!of the set of PPT states}%
 \index{state!PPT} $V_{PPT}$ (hence also
 separable states\index{state!separable})
to the volume\index{volume} of the
total set of states\index{state} goes down exponentially with the dimension of
the system\index{system}
(see Fig. \ref{fig-vol}).
This result was obtained numerically \cite{volume}, and still awaits for
analytical proof\footnote{The result could rely on the chosen measure
of the volume\index{volume!measure} \cite{Slater}.
In \cite{Karol} two different measures were
compared and produced similar results.}.  However it is compatible with the
rigorous result in \cite{infinite}
that in infinite dimension the set of
separable states\index{state!separable} is nowhere dense in the
total set of states\index{state}. Then the generic
infinite-dimensional state\index{state!entangled} is entangled.

\part{Mixed-state entanglement as a resource for quantum communication}
\label{chap-dist}
As one knows, if two distant observers (one usually calls them Alice and Bob)
share a pair of particles in
singlet state\index{state!singlet} $\psi_-$ then they can send
a quantum state\index{state!quantum}\index{quantum!state} to one another
by use of only additional two classical
bits\index{bit!classical}.
This is called  quantum
teleportation\index{teleportation}\index{quantum!teleportation} \cite{Bennett_tel}. If
the classical communication\index{communication!classical} is
free of charge (since it is much cheaper
than communication\index{communication!quantum} of
quantum\index{quantum!communication}
bits\index{bit!quantum}), one can say that a singlet pair
is a resource\index{resource} equivalent to sending
one qubit\index{qubit}. In the
following it will be
shown that mixed-state entanglement\index{entanglement!mixed-state}
can also be a resource\index{resource!for quantum communication} for
quantum communication\index{quantum!communication}\index{communication!quantum}. The
quantum communication\index{quantum!communication}\index{communication!quantum}
via mixed entangled states\index{state!mixed}%
\index{state!entangled} will require, apart from
teleportatino\index{teleportation}, the
action\index{action} called distillation\index{distillation}. 
It will be also shown that there exists a peculiar
type of entanglement\index{entanglement!types}  (bound
entanglement\index{entanglement!bound}) that
is a surprisingly
weak resource\index{resource}.

\section{Distillation of entanglement: conterfactual error correction}

Now we will attempt to describe an ingenious concept of distillation of
entanglement\index{entanglement!distillation} introduced
in Ref. \cite{Bennett_pur} and
developed in \cite{huge,Ekert} (see also \cite{Enk}).
To this end let us first briefly describe the idea of classical
and quantum  communication\index{quantum!communication}\index{communication!quantum} via noisy
channel\index{channel!noisy}\index{communication!via noisy channel}. As
one knows \cite{Cover},
the central idea of
classical information\index{information!classical} theory pioneered
by Shannon
is that one can send information\index{information} reliably and with nonzero
rate via
noisy information channel\index{channel!noisy}\index{channel!information}.
This is achieved by coding\index{coding}: the input $k$ bits\index{bit} of
information\index{information} is encoded into
a larger number of $n$ bits\index{bit}. Such a  package is sent
down the noisy channel\index{channel!noisy}.
Then, the receiver performs decoding\index{decoding}
transformation\index{transformation!decoding},
recovering the input $k$ bits\index{bit} with asymptotically
(in the limit of large $n$ and $k$) perfect
fidelity\index{fidelity!perfect}. Moreover, the
asymptotic rate of
information transmission%
\index{transmission!rate}\index{transmission!of quantum information}%
\index{information!transmission!rate}
$k\over n$ is nonzero.

In quantum domain one would like to communicate {\it quantum
states}\index{state!quantum}\index{quantum!state} instead
of classical messages. It appears,  that here the analogous scheme can be
applied \cite{Shor,Steane}. The input $k$ qubits\index{qubit} of
quantum information\index{quantum!information}\index{information!quantum}
are supplemented with additional qubits\index{qubit} in some
standard initial state\index{state},
and the total system\index{system} of $n$
qubits\index{qubit} is subjected to some
quantum
transformation\index{transformation!quantum}\index{quantum!transformation}. 
Now the package can be sent down the channel\index{channel}.
After decoding\index{decoding}
operation\index{operation!decoding}, the state\index{state} of $k$
qubits\index{qubit} is
recovered with
asymptotically perfect
fidelity\index{fidelity!asymptotically perfect} \cite{Schumacher}
(now it is {\it quantum}
fidelity\index{quantum!fidelity}\index{fidelity!quantum} -
characterising
how close is the output state\index{state!output} to the
input one) regardless of the particular
form of the state\index{state}. The discovery of the above
possibility (called
quantum
error correction\index{quantum!error correction}\index{error
correction!quantum}; we will
call it here {\it direct} error
correction\index{error correction!direct})
initiated, in particular, extensive studies of quantum error
correcting codes\index{quantum!error correcting code}\index{code!quantum
error correcting} (see
\cite{Beth} and references therein), as well
as capacities of quantum
channel\index{quantum!channel}\index{channel!quantum}%
\footnote{Capacity $Q$ of a
quantum channel\index{quantum!channel}\index{channel!quantum}
is the greatest ratio
$k\over n$ of reliable transmission\index{transmission!reliable} down
the given
channel\index{channel}.}
(see \cite{Barnum} and references therein).  A common example of
a quantum channel\index{quantum!channel}\index{channel!quantum} is
the one-qubit quantum
depolarising
channel\index{quantum!channel!depolarising}%
\index{channel!quantum!depolarizing}\index{channel!quantum!one-qubit}:
here an
input state\index{state!input} is undisturbed with
probability $p$ and subjected to a random unitary
transformation\index{transformation!random}%
\index{transformation!unitary} with probability $1-p$. It
can be described by the following
completely positive map\index{map!completely positive}
\be
\varrho\rightarrow \Lambda(\varrho)= p\varrho+(1-p){\id\over 2}\;,
\label{dep}
\ee
where $\id/2$ is the maximally mixed state\index{state!mixed!maximally} of
one
qubit\index{qubit}. This channel\index{channel} has been
thoroughly investigated \cite{Bennett_pur,huge,Bennett_cap,Bruss}.
What is important here, it was shown \cite{Bruss} that
 for $p\leq{2\over 3}$ the above method of error
 correction\index{error correction} does not
work. In classical domain it would mean that the
channel\index{channel} is useless.
Here, surprisingly,  there is a trick that allows to beat
this limit, even down to $p={1\over 3}$! The scheme that realizes it
is quite mysterious. In direct
error correction\index{error correction!direct} we deal directly
with the
systems\index{system} carrying information\index{information} to
be protected. Now, it appears
that using {\it entanglement}\index{entanglement}, one can
remove the results of  action\index{action!of noise}
 of
noise even without having the information\index{information} to
be sent. Therefore, it can be
called {\it conterfactual} error
correction\index{error correction!counterfactual}.

\begin{figure}
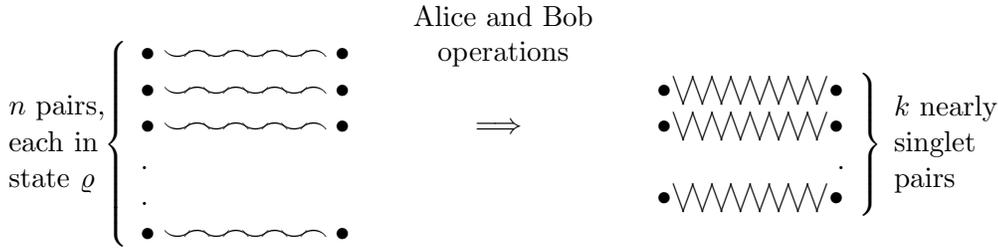

\def\weakpair{$\bullet \smile\!\!\frown\!\!\smile\!\!\frown
\!\!\smile\!\!\frown\!\!\smile\!\!\frown
\bullet$}
\def\strongpair%
{$\bullet\backslash\!/\!\backslash\!/\!\backslash\!/\!\backslash\!/%
\!\backslash\!/\!\backslash\!/\!\backslash\!/\!\backslash\!/%
\bullet$}
$
\parbox{1.8cm}{$n$  pairs, \hfil \\ each in \\state $\varrho$}
\kern-6mm
\left\{\begin{array}{l}
$\weakpair$ \\
$\weakpair$ \\
$\weakpair$ \\
.\\
.\\
$\weakpair$
\end{array} \right.
\quad
\begin{array}{c}
{\rm  Alice \  and \ Bob}\\
{\rm operations}\\
\ \\
\Longrightarrow
\ \\
\ \\
\ \\
\ \\
\ \\
\end{array}
\quad
\left.\begin{array}{r}
$\strongpair$\\
$\strongpair$\\
.\\
$\strongpair$\\
\end{array} \right\}
\hskip1mm \parbox{1.8cm}{$k$ nearly \hfil\\ \   singlet  \\ \ \ pairs}
$
\caption[Distillation of mixed-state entanglement]{Distillation of 
 mixed-state entanglement} 
\label{fig-pur}
\end{figure}

How does it work? The very idea is  not complicated. Alice (the
sender) instead of the qubits\index{qubit}
of information\index{information}, sends
to Bob particles from
entangled pairs (in state\index{state} $\psi_-$), keeping one
particle from each pair. The
pairs get disturbed by the \index{action!of channel}
action of the
channel\index{channel}, so that their state\index{state}
turns into mixture\index{mixture}\footnote{If the channel\index{channel} is
memoryless\index{channel!memoryless}, it factorises
into states\index{state} $\varrho$ of individual pairs.} that
still possesses some
residual entanglement\index{entanglement}. Now it
turns out that by local quantum
operations\index{quantum!operations!local}\index{operation!local quantum}
(including collective actions\index{action!collective}
 over all members of pairs in each lab) and
classical communication \index{communication!classical}(LQCC)
between Alice and Bob, they are able to obtain less
number of pairs in nearly maximally entangled
state\index{state!entangled!maximally} $\psi_-$
(see Fig. \ref{fig-pur}).
Such a procedure
proposed in   \cite{Bennett_pur} is
called distillation\index{distillation}. As in the case of
direct error correction\index{error correction!direct}, one
can achieve finite asymptotic
rate $k\over n$ of the distilled pairs per input pair, and
the fidelity\index{fidelity}
that now denotes similarity of the distilled pairs to the product of singlet
pairs, is asymptotically perfect.
Now, the distilled pairs can be used for
teleportation\index{teleportation}  of
quantum information\index{quantum!information}\index{information!quantum}.
The maximal possible rate achievable within the above framework is called
{\it entanglement of distillation}\index{entanglement!of distillation} of
the state\index{state} $\varrho$, and
denoted by $D(\varrho)$. Thus, if Alice and Bob share $n$ pairs
each in state\index{state} $\varrho$, they can faithfully
teleport $k=nD(\varrho)$
qubits\index{qubit}.

As we have mentioned, the error correction\index{error correction} stage
and the transmission stage\index{transmission} are here
separated in time\index{time}; the error correction\index{error correction} can
be performed even before the information\index{information} to be
sent was produced.
Using terminology of \cite{clon} one can say, that Alice and Bob operate
on {\it potentialities} (entangled pair represents a potential communication%
\index{communication!potential})
and correct potential error, so that, when the {\it actual} information
is coming, it can be teleported without any
additional action\index{action}.

The above scheme is not only mysterious. It is also much more powerful from
the direct method. In next section we describe a distillation
protocol\index{protocol!of distillation}\index{distillation!protocol}, that
allows to send quantum information\index{quantum!information}\index{information!quantum}
reliably via the
channel\index{channel} with $p>1/3$.
A general question  is: where are the limits
of distillation\index{distillation}? As we have seen,
the basic action refer mixed bipartite
state\index{state!mixed}\index{state!bipartite}, so that instead
of saying about
channels\index{channel}, we can concentrate
on bipartite states\index{state!bipartite}. The question can be
formulated: which states\index{state} $\varrho$ can be distilled by the
most general LQCC actions\index{action!LQCC}?
Here, saying that a {\it state\index{state} $\varrho$}
can be distilled, we mean that Alice and Bob can obtain singlets from
the initial state\index{state} $\varrho^{\otimes n}$ of $n$ pairs
(thus we will work with memoryless channel\index{channel!memoryless}).

One can easily see, that separable states\index{state!separable} cannot
be distilled: they contain
no entanglement\index{entanglement}, so it
is impossible to convert them into entangled  ones
by LQCC
operations\index{operations!LQCC}. Then the final
form of our question is:
Can all  entangled states\index{state!entangled} can be
distilled? Before answer the to this question
was provided, the default was ``yes'', and  the problem was: how to prove it.
Now, one knows that the answer is ``no'', so  that the structure
of entanglement\index{entanglement!of bipartite states} of bipartite
states\index{state!bipartite} is much more puzzling than one could
suspect.

Finally, one should mention that for pure states\index{state!pure} the problem of conversion
into singlet pairs is solved. Here there is no surprise: all entangled pure
states\index{state!pure} can be distilled \cite{conc} (see also \cite{Nielsen} in this context).
What is especially important, this distillation\index{distillation}
can be made reversibly: from
the obtained singlet pairs, we can recover (asymptotically) the same number
of input pairs \cite{conc}. As we will see, for mixed
states\index{state!mixed} it
is not the case.

\section{Distillation of two-qubit states}
\label{sec-two-qubits}

In this section we will describe historically the first distillation protocol%
\index{distillation!protocol}\index{protocol!of distillation}
for two-qubit states\index{state!two-qubit} devised by 
Bennett, Brassard, Popescu, Schumacher, Smolin and Wootters (BBPSSW)  
\cite{Bennett_pur}. Then
we will show that a more general
protocol\index{protocol} can distill any entangled two-qubit
state\index{state!two-qubit}\footnote{In this contribution we restrict ourselves
to distillation\index{distillation!protocol}
by means of perfect operations\index{operations}. The more realistic case of imperfections
of the quantum
operations\index{quantum!operations}\index{operation!quantum} performed by
Alice and Bob is
considered in \cite{Briegel}}.

\subsection{BBPSSW distillation protocol}
\label{sec-Bennett}
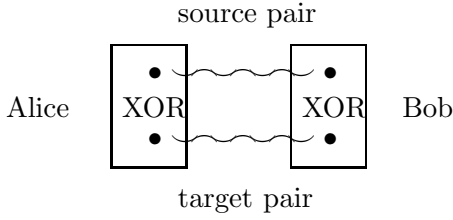
\begin{figure}
\def\para{
$\bullet \smile\!\!\frown\!\!\smile\!\!\frown
\!\!\smile\!\!\frown\!\!\smile
\bullet$}
\parbox{200pt}{
\begin{picture}(200,100)
\put(95,82){source pair}
\put(80,35){\para}
\put(80,60){\para}
\put(30,46){Alice}
\put(74,46){XOR}
\put(70,27){\framebox(28,46){}}
\put(142,46){XOR}
\put(138,27){\framebox(28,46){}}
\put(180,46){Bob}
\put(95,12){target pair}
\end{picture}}
\caption[Bilateral quantum XOR transformation]
{Bilateral quantum XOR operation\index{operation!XOR}}
\label{fig-bxor}
\end{figure}

The BBPSSW distillation\index{distillation}.
 protocol\index{protocol!BBPSSW}
still remains the most transparent example of
distillation\index{distillation}.
It works for the two-qubit states\index{state!two-qubit} $\varrho$ with fully entangled
fraction\index{fraction!fully entangled}
satisfying $\fcal>1/2$. Such  states\index{state} are equivalent to the ones with
$F>1/2$, so that we can restrict ourselves to the
latter states\index{state}.
Hence we assume that Alice and Bob share initially a huge number of
pairs, each in the same state\index{state} $\varrho$ with $F>1/2$, so that
the total state\index{state} is $\varrho^{\otimes n}$. Now, they  aim to obtain
a smaller number of pairs of higher
singlet fraction\index{fraction!singlet} $F$. To this end
they iterate the following steps:
\bei
\item[1.] They take two pairs, apply to each of them  $U\otimes U^*$
twirling\index{twirling!$U\otimes U^*$} i.e. random unitary
transformation\index{transformation!unitary}\index{transformation!random}
of the form  $U\otimes U^*$\index{transformation!$U\otimes U^*$}
(Alice picks at random a
transformation\index{transformation} $U$, applies it, and communicate Bob,
which transformation\index{transformation} she
chose; then he applies $U^*$ to his particle).
Thus one has transformation\index{transformation} from
two copies of $\varrho$ to
two copies of isotropic state\index{state!isotropic} $\varrho_F$
with unchanged $F$:
\be
\varrho\otimes\varrho \rightarrow \varrho_F\otimes \varrho_F\;.
\ee
\item[2.] Each party performs unitary
transformation\index{transformation!unitary!XOR} XOR%
\footnote{Quantum XOR gate\index{quantum!gate!XOR}\index{gate!quantum XOR}
is the
most common quantum\index{quantum!gate!two-qubit}
two-qubit gate\index{gate!two-qubit}\index{gate!quantum}
introduced in \cite{d1}.}
on their members of pairs (see Fig. \ref{fig-bxor}).
The transformation is given by
\be
U_{XOR}|a\ra|b\ra=|a\ra|(a+b)\mathop{\rm mod} 2\ra
\ee
(the first qubit\index{qubit} is called source, the second one - target).
They obtain some complicated state\index{state} $\tilde\varrho$ of two pairs.
\item[3.] The pair of target qubits\index{qubit!target} is locally measured
in the
basis\index{basis} $|0\ra,|1\ra$ and it is discarded.
If the results agree (success), the source pair is kept
and has a greater singlet fraction\index{fraction!singlet}.
Otherwise (failure) the source pair is
discarded, too.
\eei
If the results in step 3 agreed the final
state\index{state} $\varrho'$ of the kept source pair
can be calculated from the formula
\be
\varrho'=\trace_{\hcal_t}(P_t\otimes \id_s \tilde \varrho P_t\otimes \id_s)\;,
\ee
where the partial trace is performed over the
Hilbert space\index{space!Hilbert} $\hcal(t)$
of the target pair, $\id_s$ is identity on the
space\index{space} of source pair
(because it was not measured), while $P_t=|00\ra\la00|+|11\ra\la11|$
acts on target pair space\index{space} and
corresponds to the case ``results agree''.

Subsequently, one can calculate the
singlet fraction\index{fraction!singlet} of the survived pair
as a function of the singlet fraction\index{fraction!singlet} of
the two initial ones obtaining
\be
F'(F)={F^2+{1\over 9}(1-F)^2\over F^2+{2\over 3}F(1-F)+{5\over 9}(1-F)^2}\;.
\ee
Since the function $F(F')$ is continuous, $F'(F)>F$ for $F>{1\over 2}$ and
$F'(1)=1$, we obtain that iterating the procedure Alice and Bob can obtain
state\index{state} with arbitrarily high  $F$. Of course, the larger
$F$ is required
the more pairs must be sacrificed, and the less the probability $p$ of the
success is. Thus if they start with some $F_{in}$ and would like to
end up with some higher $F_{out}$, the number of final pairs will be on average
$k=np/{2^l}$, where $l$ and $p$ depend on $F_{in}$ and $F_{out}$, and denote
respectively the number of iterations of the function $F'(F)$ required to
reach $F_{out}$ starting from $F_{in}$, and the probability of the string of
$l$ successful operations\index{operations}.

The above method allows to obtain arbitrarily high $F$, but the asymptotic
rate is zero.
However, if $F$ is high enough that $1-S>0$  where $S$ is
von Neumann entropy\index{entropy!von Neumann} of the
state\index{state} $\varrho$ , then there exists a
protocol (so called hashing),
\index{protocol!hashing}
that gives nonzero rate
\cite{huge}. We will not describe it here, but we note that
for any state\index{state} with $F>{1\over 2}$ Alice and Bob can
start by the recurrence
method to obtain $1-S>0$, and then apply the
hashing protocol\index{protocol!hashing}. This gives
nonzero rate for any state\index{state} with $F>1/2$. This means that quantum
information\index{quantum!information}\index{information!quantum} can
be transmitted via depolarising
channel\index{channel!depolarising} (\ref{dep})
if only $p>1/3$. Indeed, one can check that if Alice send one of
the particles from a pair in a state\index{state} $\psi_+$ via the
channel\index{channel} to Bob,
then the final state\index{state} shared by them will be
isotropic one  with $F>1/2$. Repeating this process, Alice and Bob can
obtain many such pairs. Then distillation\index{distillation} will allow to use them for
asymptotically faithful quantum
communication\index{quantum!communication}\index{communication!quantum}.

\subsection{All entangled two-qubit states are distillable}
\label{sec-pur}

As it was mentioned in Sect. \ref{examples}, there exist entangled two-qubit
states\index{state!two-qubit} with $\fcal<1/2$, so that no product
unitary transformation\index{transformation!unitary}
can produce $F>1/2$. Thus the BBPSSW
protocol\index{protocol!BBPSSW}  does not apply
to all entangled two-qubit states\index{state!two-qubit}%
\index{state!entangled}. Below we will show that, nevertheless,
all such states\index{state!distillable} are
distillable \cite{pur}. The problem was possible
to solve mainly due to the  characterisation of the
entangled states\index{state!entangled} as discussed
in Sect. \ref{sec-ent-pos}.

Since we are not interested in the value of asymptotic rate, it suffices to
show that starting with pairs in an
entangled state\index{state!entangled}, Alice and Bob
are able to obtain a fraction of them  in a new
state\index{state} with $F>1/2$
(then the BBPSSW
protocol\index{protocol!BBPSSW} will do the job).  Our
main tool will
be the so called filtering\index{filtering}\index{filtering!operation}
operation\index{operation!filtering} \cite{Gisin,conc} that
involves {\it generalised}
measurement\index{measurement!generalized} performed by one
of the parties (say, Alice) on individual pairs.
It will consists of two
outcomes $\{1,2\}$, associated with
operators\index{operator} $W_1$ and $W_2$ satisfying
\be
W_1^\dagger W_1+ W_2^\dagger W_2=\id_A
\label{tracepr}
\ee
($\id_{A(B)}$ denote  identity on Alice (Bob) system\index{system}).
After such measurement\index{measurement}, the
state\index{state} becomes
\be
\varrho\rightarrow
{1\over p_i} W_i\otimes \id_B\,\varrho\, W_i^\dagger\otimes \id_B ,\quad i=1,2
\label{wynik}
\ee
with probability  $p_i=\trace
(W_i\varrho W_i^\dagger)$.
The condition (\ref{tracepr}) ensures  $p_1+p_2=1$.
Now Alice will be interested only in one outcome (say, 1). If it occurs,
Alice and Bob keep the pair, otherwise they discard it (this requires
communication\index{communication} from Alice to Bob). Then
we are only interested
in the operator\index{operator} $W_1$. If its norm\index{norm} does not
exceed $1$, one can always find
suitable $W_2$, so that the condition (\ref{tracepr}) is satisfied. Now, since
either the form of the final state\index{state} (\ref{wynik})
or the fact whether $p_1$ is zero or not, do not depend on the positive
factor multiplying $W_1$, we are free to consider completely arbitrary
filtering
operators\index{operator!filtering}\index{filtering!operator} $W_1$.
In conclusion,  for any entangled state\index{state!entangled} $\varrho$
we must find such
 an operator\index{operator}
$W$ that the  state\index{state} resulting from
filtering\index{filtering}
$W\otimes \id\,\varrho\, W^\dagger\otimes \id /\trace
(W\otimes \id\,\varrho\, W^\dagger\otimes \id)$ has $F>1/2$.
Consider then an arbitrary two-qubit entangled
state\index{state!entangled}\index{state!two-qubit} $\varrho$.
From the Theorem \ref{2x2} we know that $\varrho^{T_B}$ is not a positive
operator\index{operator!positive}, hence there exists a
vector\index{vector} $\psi$ for which
\be
\la \psi|\varrho^{T_B}|\psi\ra<0\;.
\label{neg}
\ee
Now let us note that any vector\index{vector} $\phi$ of $d\otimes d$
system\index{system!$d\otimes d$} can be written
as $\phi=A_\phi\otimes \id \psi_+$, where $A_\phi$ is
some operator\index{operator}.
Indeed, write $\phi$ in  product basis\index{basis!product}:
$\phi=\sum_{i,j=1}^da_{ij}|i\ra\otimes|j\ra$. Then the
matrix elements\index{matrix!elements} of the
operator\index{operator} $A_\phi$ are given by $\la i|A_\phi|j\ra=\sqrt d
a_{ij}$ (in our
case $d=2$). Therefore the formula (\ref{neg}) can be rewritten the form
\be
\trace \bigl[(A_\psi^\dagger \otimes \id\, \varrho\,
A_\psi\otimes \id)^{T_B}\,P_+\bigr]<0\;.
\ee
Using identity $\trace A^{T_B} B=\trace AB^{T_B}$  valid for any
operators\index{operator}
$A,B$ and the fact that $P_+^{T_B}={1\over d}V$ (where $V$ is the
flip operator\index{operator!flip},
see Sect. \ref{examples}) we obtain
\be
\trace \bigl[(A_\psi^\dagger \otimes \id \,\varrho\, A_\psi\otimes \id)\,
V\bigr]<0\;.
\label{in-v}
\ee
We conclude that $ A_\psi^\dagger \otimes \id\,\varrho\, A_\psi\otimes \id$
cannot be equal to
null operator\index{operator!null}, hence we can
consider the following state\index{state}
\[
\tilde\varrho=
{A_\psi^\dagger \otimes \id\,\varrho\, A_\psi\otimes \id
\over \trace(A_\psi^\dagger \otimes\id\, \varrho\, A_\psi\otimes \id)}\;.
\]
Now it is clear that
the role of filter\index{filter} $W$ will be played
by operator\index{operator} $A_\psi^\dagger$.
We will show that $\la\psi_-|\tilde\varrho|\psi_-\ra>1/2$ where
$\psi_-=(|01\ra-|10\ra)/\sqrt2$. Then a  suitable Alice's unitary
transformation\index{transformation!unitary} can
convert $\tilde\varrho$ into a
state\index{state} $\varrho'$
with $F>1/2$.

From the inequality (\ref{in-v}) one obtains
\be
\trace\tilde \varrho V<0\;.
\label{in-v2}
\ee

If we use product
basis\index{basis!product} $|1\ra=|00\ra,|2\ra=|01\ra,
|3\ra=|10\ra,|4\ra=|11\ra$
inequality (\ref{in-v2}) writes as
\be
\tilde\varrho_{11}+\tilde\varrho_{44}+
\tilde\varrho_{23}+\tilde\varrho_{32}<0\;.
\ee
The above inequality, together with the trace condition
$\trace\tilde\varrho=\sum_i\tilde\varrho_{ii}=1$ gives
\be
\la\psi_-|\tilde \varrho|\psi_-\ra={1\over 2}
(\tilde\varrho_{22}+\tilde\varrho_{33}
-\tilde\varrho_{23}-\tilde\varrho_{32})>{1\over 2}\;.
\ee

To summarise, given a large supply of pairs, each in
entangled state\index{state!entangled} $\varrho$
Alice and Bob can distill maximally entangled pairs in the following way.
First
Alice applies filtering\index{filtering} determined by the
operator\index{operator} $W=A_\psi^\dagger$
described
above. Then Alice and Bob  obtain on average a supply of $np$ survived pairs
in the state\index{state} $\tilde
\varrho$ (here $p=\trace W\otimes \id\, \varrho\, W^\dagger\otimes \id$ is
the probability that the outcome of Alice measurement\index{measurement}
will be the one  associated with the operator\index{operator} $W^\dagger$).
Now Alice applies  an operation\index{operation}
$i\sigma_y$ to obtain  the state\index{state} with  $F>1/2$. Then they
can use the BBPSSW protocol\index{protocol!BBPSSW} to
distill maximally
entangled pairs useful
for quantum communication%
\index{quantum!communication}\index{communication!quantum}. Note that
we assumed that Alice and Bob know the
initial state\index{state} of the pairs. It can be shown that
if they do not know, they
still can do the job (in the two-qubit case)
sacrificing $\sqrt n$  pairs to estimate the
 state\index{state} \cite{Pawel}.

The above protocol\index{protocol} can easily by shown
to work in $2\otimes 3$ case.
The protocol\index{protocol} can be also fruitfully applied for
 the system $2\otimes n$\index{system!$2\otimes n$} if
 the state is NPT\index{state!NPT}
 \cite{Lewen_Cir}.

\section{Examples}
\label{sec-dist-examples}

Consider the state\index{state} (\ref{przyklad})
from Sect. \ref{examples}
$\varrho=p|\psi_-\ra\la\psi_-|+(1-p)|00\ra\la00|$. It is entangled for
all $p>0$. In matrix notation we have
\be
\varrho=
\left[
\ba{cccc}
1-p &0&0&0\\
0 &p \over 2 &-{p\over 2} &0\\
0 &-{p \over 2} &{p\over 2} &0\\
0&0&0&0\\
\ea
\right]\;,
\quad
\varrho^{T_B}=
\left[
\ba{cccc}
1-p &0&0&-{p\over 2}\\
0 &p \over 2 &0 &0\\
0 &0 &{p\over 2} &0\\
-{p\over 2}&0&0&0\\
\ea
\right]\;.
\ee
The negative eigenvalue\index{eigenvalues!negative} of
$\varrho^{T_B}$ is $\lambda_-={1\over 2}
\left(1-p-\sqrt{(1-p)^2+p^2}\right)$ with the corresponding (unnormalised)
eigenvector\index{eigenvectors} $\psi=\lambda_-|00\ra-{p\over 2}|11\ra$,
hence we can take the filter\index{filter} of the form 
$W={\rm diag}[\lambda_-,-{p\over 2}]$. The new
state\index{state} $\tilde\varrho$
resulting from filtering\index{filtering} is of the form
\be
\tilde\varrho=
{1\over N}\left[
\ba{cccc}
\lambda_-^2(1-p) &0&0&0\\
0 &p^3 \over 8 &{p^2\over 4}\lambda_- &0\\
0 &{p^2\over 4}\lambda_- &{p\over 2}\lambda_-^2 &0\\
0&0&0&0\\
\ea
\right]\;.
\ee
where $N=\lambda_-^2(1-p)+{p^2/8}+\lambda_-^2 p/2$.
Now the overlap with $\psi_-$ given by $\la\psi_-|\tilde\varrho|\psi_-\ra=
(p^3/8+\lambda^2p/2-\lambda p^2/2)/N$ is greater than $1/2$ if only $p>0$.
The new state\index{state} can be distilled by BBPSSW
protocol\index{protocol!BBPSSW}.

Below  we will prove that some  states\index{state!distillable} of
higher dimensional systems\index{system!higher-dimensional}
are distillable. We will do it by showing that some
LQCC operation\index{operation!LQCC}
can convert them (possibly with some probability) into entangled
two-qubit state\index{state!entangled}\index{state!two-qubit}.
\subsection{Distillation of isotropic state
for $d\otimes d$ system\index{system!$d\otimes d$}.}
For $F>1/d$ isotropic state\index{state!isotropic} can be distilled
\cite{xor,nmr}.
If {\it both} Alice and Bob apply the
projector\index{projector} $P=|0\ra\la0|+|1\ra\la1|$
where $|0\ra$, $|1\ra$ are vectors\index{vector} from the local
basis\index{basis!local}, then the
isotropic state\index{state!isotropic}  will be
converted into two-qubit isotropic state\index{state!isotropic!two-qubit}.
(Note that
the projectors\index{projector} play the
role of filters\index{filter}; also, the success of the
filtering\index{filtering} is
if both Alice and Bob obtain outcome corresponding to $P$). Now, if
the initial state\index{state} satisfied $F>1/d$ then the final one, as
a two qubit one,
will have $F>1/2$. Thus, it is entangled hence can be distilled.

\subsection{Distillation and reduction criterion}
Any state\index{state} $\varrho$ of $d\otimes d$
system\index{system!$d\otimes d$}
that violates reduction criterion\index{criterion!reduction}
(see Sect. \ref{examples})
can be distilled \cite{xor}. Indeed, take the vector\index{vector} $\psi$ for which
$\la\psi|\varrho_A\otimes \id-\varrho|\psi\ra<0$. Now, it is easy to see that
applying the
filter\index{filter} $W$ given by $\psi=W\otimes \id\psi_+$, one obtains
state\index{state} with $F>{1\over d}$. Now, the
random $U\otimes U^*$ transformations
\index{transformation!$U\otimes U^*$}
\index{transformation!random}
will convert it into isotropic state\index{state!isotropic} with
the same $F$. As shown above the
latter one is distillable.

\section{Bound entanglement}
\label{sec-bound}
In the light of the result for two qubits\index{qubit}, one
naturally expected
that any entangled state can be distilled\index{state!entangled}. It
was a great surprise when it
appeared that it is not the case. In \cite{bound} it has been shown that
there exist entangled states\index{state!entangled}, that cannot
be distilled.
The following  theorem provides necessary and sufficient condition%
\index{condition!for distillability!necessary and sufficient} for
distillability of mixed states\index{state!mixed} \cite{bound}.
\begin{theorem}
A state\index{state!distillable} $\varrho$ is distillable if
and only if for some two-di\-men\-sional
projectors\index{projector!two-dimensional} $P,Q$ and
for some number $n$, the state\index{state} $P\otimes Q
\varrho^{\otimes n} P\otimes Q$ is entangled.
\label{dist-char}
\end{theorem}
\textbf{Remarks.}
(1) Note that the state\index{state}
$P\otimes Q \varrho^{\otimes n} P\otimes Q$ is
effectively two-qubit one as its
support\index{support} is contained in the
$\Complex^2\otimes \Complex^2$ subspace\index{subspace} determined by
the projectors\index{projector} $P,Q$. This means that
the distillable entanglement\index{entanglement!distillable}
is {\it two-qubit} entanglement\index{entanglement!two-qubit}.
(2) One can see that the theorem is compatible with the fact \cite{conc}
that any {\it pure} state\index{state!pure} can be distilled.

As a consequence of this theorem we obtain the following one \cite{bound}:

\begin{theorem}
A PPT state\index{state!PPT} cannot be distilled.
\label{PPT-bound}
\end{theorem}

\textbf{Proof.} We will give here a proof independent of the Theorem
\ref{dist-char}. As a matter of fact, we will show  that
the set of PPT states\index{state!PPT}
is (i) invariant under LQCC  operations\index{operation!LQCC}
\cite{bound} and   (ii)
it is bounded away from maximally
entangled state\index{state!entangled!maximally} \cite{Dagstuhl,Rains}.
Then, since
$(\varrho^{\otimes n})^{T_B}=(\varrho^{T_B})^{\otimes n}$ we obtain the proof
of the theorem.
To prove (i) note that
any LQCC operation\index{operation!LQCC} can be written as \cite{Plenio}
\be
\varrho\rightarrow
\varrho'={1\over p}\sum_i A_i\otimes B_i \varrho A^\dagger_i
\otimes B^\dagger_i\;,
\ee
where $p$ is a normalisation
constant
 interpreted as  probability  of
realization of the
operation\index{operation}, and  the map\index{map} $\varrho\rightarrow
\sum_i A_i\otimes B_i \varrho A^\dagger_i \otimes B^\dagger_i$ does not
increase trace
(this ensures $p\leq1$). Suppose now that $\varrho$ is PPT, i.e.
$\varrho^{T_B}\geq0$ and examine partial transposition of the
state\index{state} $\varrho'$.
We will use the following property of partial transpose
\be
(A\otimes B\varrho C\otimes D)^{T_B}=A\otimes D^T\varrho^{T_B}
C\otimes B^T
\label{eq}
\ee
for any operators\index{operator} $A,B,C,D$ and $\varrho$.
Then we obtain
\be
(\varrho')^{T_B}= \sum_i A_i\otimes (B_i^\dagger)^T \varrho^{T_B}
A_i\otimes (B_i)^T\;.
\ee
Thus $(\varrho')^{T_B}$ is a result of
action\index{action!of completely positive map} of some
completely positive map\index{map!completely positive}
on operator\index{operator} $\varrho^{T_B}$ that by assumption is positive. Then also the
operator\index{operator!positive} $(\varrho')^{T_B}$ must be positive.  Thus
LQCC map\index{map!LQCC} do not move
outside the set of PPT states\index{state!PPT}.

To prove (ii) let us now show that PPT states\index{state!PPT} can never have
high singlet fraction\index{fraction!singlet} $F$.
Consider a PPT state\index{state!PPT} $\varrho$ of a $d\otimes d$
system\index{system!$d\otimes d$}. We obtain
\be
\trace \varrho P_+= \trace \varrho^{T_B} P_+^{T_B}\;.
\ee
Now, it is easy to check that $P_+={1\over d} V$, where $V$ is the flip
operator\index{operator!positive} described in section \ref{examples}.
Note that $V$ is Hermitian and
has eigenvalues\index{eigenvalues} $\pm1$.
Since $\varrho$ is PPT then 
$\tilde\varrho=\varrho^{T_B}$ is a legitimate state\index{state}, and
the above expression
can be rewritten in terms of mean  value of
the observable\index{mean value!of observable}\index{observable}
$V$
\be
\trace \varrho P_+={1\over d}\trace \tilde\varrho V\;.
\ee
The mean value of dichotomic
observable\index{observable!dichotomic}\index{mean value!of observable}
cannot exceed $1$ so that we obtain
\be
F(\varrho)\leq {1\over d}\;.
\ee
Thus the maximal possible singlet fraction\index{fraction!singlet}
that can be attained by PPT states\index{state!PPT} is
the one that can be obtained without any prior
entanglement\index{entanglement} between the
parties. Indeed, a {\it product} state\index{state!product} $|00\ra$ has
singlet fraction\index{fraction!singlet} $1/d$
(if
it belongs to the Hilbert space\index{space!Hilbert}
$C^d\otimes C^d$). Consequently, for however
large amount of PPT pairs, even a single two-qubit pair with $F>1/2$ cannot
be obtained by LQCC actions\index{action!LQCC}. \blacksquare

Now, one can appreciate the results presented in the first part of
this contribution. From Sect. \ref{sec-ent-pos}  we know that there
exist entangled states\index{state!entangled} that are
PPT\index{state!PPT}. So far, the question of whether
there exist entangled states\index{state!entangled} that
are  PPT\index{state!PPT} was merely a technical one.
At present, since the above theorem implies that
PPT states\index{state!PPT} are
non-distillable, we can draw a remarkable conclusion: there exist
non-distillable entangled states\index{state!entangled!non-distillable}.
Since in the process of distillation%
\index{distillation!process}
no entanglement\index{entanglement} can be liberated
to the useful singlet form,
they have been called {\it bound entangled}. Thus there exist at least two
{\it qualitatively different} types of
entanglement\index{entanglement!types}:
apart from  the  {\it free} entanglement\index{entanglement!free} that
can be distilled, there is a
{\it bound} one that cannot be distilled and seems to be completely useless for
quantum communication\index{quantum!communication}\index{communication!quantum}. This
{\it discontinuity} of the structure of
entanglement of mixed states\index{state!mixed}%
\index{entanglement!mixed-state} was
considered to be possible
for multipartite systems\index{system!multipartite}, but it
was completely surprising for bipartite
systems\index{system!bipartite}. It should be emphasised here, that the
BE states\index{state!bound entangled (BE)} are not atypical
in the set of all possible states\index{state}: as we have
mentioned in Sect. \ref{vol}
the volume of the PPT entangled states%
\index{volume!of the set of PPT entangled states}%
\index{state!PPT entangled} is nonzero.
One of the main consequence of existence of BE
is revealing  a transparent form
of {\it irreversibility}\index{irreversibility!in entanglement processing}
in
entanglement processing\index{entanglement!processing}. If Alice and
Bob share pairs in pure state\index{state!pure},
then to produce BE state\index{state!bound entangled (BE)} they need some prior
entanglement\index{entanglement}. However once they
produced the BE states\index{state!bound entangled (BE)}, they would not be able to
recover the pure
entanglement\index{entanglement!pure} back from
them. It is {\it entirely} lost. This is a
{\it qualitative}
irreversibility\index{irreversibility!qualitative}
that is  probably a
source of
the {\it quantitative} irreversibility
\index{irreversibility!quantitative}\cite{Bennett_pur,huge}
due to the fact that we need more pure
entanglement\index{entanglement!pure} to produce some mixed
states\index{state!mixed} than we can
then distill back from them \cite{Rains1,VP98}%
\footnote{%
The quantitative irreversibility
\index{irreversibiliyt!quantitative} was rigorously proved in \cite{irrev}.
There is still no fully rigorous proof for qualitative one 
(see \cite{irrev}).}.

To analyse the phenomenon of bound
entanglement\index{entanglement!bound}, one needs as
many examples of BE states\index{state!bound entangled (BE)} as possible. 
Then there is a very exciting
physical motivation for search for
PPT entangled states\index{state!PPT entangled}. In Sect.
\ref{PPT-ent} we discussed different methods of the search. As a result
we have a couple of examples of BE states\index{state!bound entangled (BE)}
obtained
via the separability criterion\index{criterion!separability} given by Theorem \ref{thPawel}, from
the mathematical
literature on non-decomposable maps\index{map!positive!non-decomposable},
and via unextendible
product bases\index{basis!product!unextendible (UPB)} method.

The examples produced via
UPB\index{basis!product!unextendible (UPB)} are extremely interesting from the physical
point of view. It is because UPB\index{basis!product!unextendible (UPB)} is not only a
mathematical object: as shown in \cite{UPB1} it produces a very curious
physical effect
\cite{Bennett99} called ``nonlocality
without entanglement''\index{nonlocality!without entanglement}.  Namely,
suppose that Alice and Bob share
a pair in one of the states\index{state} from the
UPB\index{basis!product!unextendible (UPB)}, but they do not know  which one it is.
It appears that by LQCC operations\index{operations!LQCC} (with
finite resources\index{resource!finite}), they are not
able to read the identity of the state\index{state}.
However, if the particles were together, then, since the
states\index{state}
are orthogonal, they can be perfectly distinguished from each other. Thus we
have a highly
{\it non classical}  effect produced by ensemble of {\it separable}
states\index{state!separable}.
On the other hand, the BE state\index{state!bound entangled (BE)} associated with the
given UPB\index{basis!product!unextendible (UPB)}
(the uniform state\index{state} on the complementary
subspace\index{subspace!complementary}, see
(\ref{UPB-bound})) presents opposite features: it {\it is} entangled but,
since its entanglement is bound\index{entanglement!bound}, it
ceases to behave quantumly. Moreover in
both situations we have a kind of
irreversibility\index{irreversibility}. As it
was mentioned, BE states\index{state!bound entangled (BE)} are reflection of the
formation--distillation
irreversibility\index{irreversibility!formation-distillation}: to create
them by LQCC from singlet
pairs, Alice and Bob need to non-zero amount of the latter. However, once they
were created, there is no way to distill singlets out of them.
 On the other hand, UPB\index{basis!product!unextendible (UPB)}
 exhibits preparation--measurement\index{measurement}
irreversibility\index{irreversibility!preparation-measurement}: any of
the states\index{state} belonging to UPB\index{basis!product!unextendible (UPB)} can be prepared by LQCC
operations\index{operations!LQCC}, but once Alice and
Bob forgot the identity of the state\index{state},
they cannot recover it by LQCC. This surprising connection between
some BE states\index{state!bound entangled (BE)} and bases\index{basis} that are not
distinguishable by LQCC implies many
interesting questions concerning future unification of our
knowledge about nature of
quantum information\index{quantum!information}\index{information!quantum}.

Finally, we will mention about the result concerning {\it
rank}\index{rank!of BE states} of the BE state\index{state!bound entangled (BE)}.
In numerical analysis of BE states\index{state!bound entangled (BE)} (especially their
tensor products\index{product!tensor}) it is
very convenient to have examples with low rank\index{rank!low}.
However,
in   \cite{PSTT}   the following bound on the
rank\index{rank!of BE states} of
the BE state\index{state!bound entangled (BE)} $\varrho$ was derived
\be
R(\varrho)\geq \max \{R(\varrho_A),R(\varrho_B)\}.
\ee
(recall that $R(\varrho)$ denotes the rank \index{rank} of $\varrho$). 
Note that the above inequality is nothing but the entropic
inequality\index{inequalities!entropic}
(\ref{entrop-ineq}) with entropy (\ref{rank}). Thus it appears 
that the latter inequality is not only necessary condition for
separability\index{condition!for separability!necessary}, but also
for non-distillability\index{condition!for non-distillability}.
The proof bases on the fact \cite{xor} that any state\index{state}
violating reduction
criterion\index{criterion!reduction} (see
Sect. \ref{examples} and \ref{sec-dist-examples}) can be
distilled. It can be shown, that
if a state\index{state}
 violate the above equation, then it must also violate reduction
criterion\index{criterion!reduction}, hence can be distilled.  Then it
follows that there do not exist BE
states\index{state!bound entangled (BE)} of rank\index{rank} two \cite{PSTT}.
Indeed, if it existed,
then its local ranks\index{rank!local}
must have not exceed two. Hence the total state\index{state} would be effectively
two-qubit one. However, from Sect. \ref{sec-pur} we know that two-qubit bound
entangled states\index{state!entangled} do
not exist.

\section{Do there exist bound entangled NPT states?}
\label{sec-search}

So far we have considered BE states\index{state!bound entangled (BE)} due to the
Theorem  \ref{PPT-bound} which says that NPT
condition\index{condition!NPT} is
necessary for
distillability\index{condition!for distillability!necessary}.
\index{condition!for distillability!necessary}
As mentioned in Sect. \ref{sec-pur}, for $2\otimes n$
systems\index{system!$2\otimes n$} all
NPT states\index{state!NPT} can be distilled \cite{Lewen_Cir}, hence the
condition is also 
sufficient\index{condition!for distillability!sufficient} in this case.
However, it is not known whether
it is sufficient in general. 
The necessary and sufficient
condition\index{condition!for distillability!necessary and sufficient} is
given by the Theorem
\ref{dist-char}. To find if the condition is equivalent to PPT one, it
must
be 
determined if there exists   an  NPT state\index{state!NPT},  such that,
nevertheless, for any number of copies $n$,
the state\index{state} $\varrho^{\otimes n}$
will not have an entangled two-qubit ``substate'' (i.e. the state\index{state}
$P\otimes Q\varrho^{\otimes n} P\otimes Q$). In \cite{xor} it was pointed out
that one can reduce the problem by means of the following observation.
\begin{proposition}
The following statements are equivalent:
\bei
\item[1.] Any NPT state\index{state!NPT} is
distillable\index{state!distillable}.
\item[2.] Any entangled Werner
state\index{state!Werner}\index{state!entangled} (eq. \ref{Werner-state}) is
distillable\index{state!distillable}.
\eei
\end{proposition}
\textbf{Proof.} The proof of the implication $(1)\Rightarrow (2)$ is immediate,
as Werner states\index{state!Werner} are
entangled\index{state!entangled} if and only if they
are NPT\index{state!NPT}. Then if we can
distill any NPT state\index{state!NPT}, then also
Werner entangled states\index{state!Werner}\index{state!entangled} are
distillable\index{state!distillable}.
To obtain  $(2)\Rightarrow (1)$ note that the reasoning
of Sect. \ref{sec-pur}, 
from formula (\ref{neg}) to (\ref{in-v2}), is insensitive to the
dimension $d$ of the problem. Consequently,
from any NPT state\index{state!NPT} 
a suitable filtering\index{filtering} 
produces 
a state $\tilde\varrho$ 
satisfying $\trace \tilde\varrho V<0$. 
As mentioned in Sect. \ref{examples}, the parameter $\trace\varrho V$ 
is invariant under $U\otimes U$ twirling\index{twirling!$U\otimes U$}, so
that applying the latter
(which is LQCC operation\index{operations!LQCC}) Alice and
Bob  obtain  Werner
state\index{state!Werner} $\varrho_W$ satisfying
$\trace\varrho_W V<0$.
Thus any NPT state\index{state!NPT} can
be converted by means
of LQCC operations\index{operations!LQCC} into
entangled Werner state\index{state!Werner}\index{state!entangled}, which
completes the proof.

\begin{figure}
\ \vskip1.5cm
\parbox{200pt}{
\begin{picture}(200,150)
\put(28,175){(a)}
\put(150,100){\oval(240,120)}    
\put(150,40){\line(0,1){120}}    
\put(80,170){\bf PPT}
\put(200,170){\bf NPT}
\put(61,95){\parbox{2.6cm}{\large \bf separable \\ \centerline{states \ \ \ \ }}}
\put(181,102){\parbox{2.6cm}{\large \bf \phantom{l} \ \ \ \ free\\ 
entangled\\ \centerline{states \ \ }}}
\end{picture}}

\vskip1cm

\parbox{200pt}{
\begin{picture}(200,150)
\put(28,175){(b)}
\put(150,100){\oval(240,120)}    
\put(150,40){\line(0,1){85}}    
\put(150,150){\line(0,1){10}}    
\put(92,96){\oval(80,50)}
\put(80,170){\bf PPT}
\put(200,170){\bf NPT}
\put(61,95){\ \parbox{2.4cm}{\large \bf separable \\ \centerline{states\ \ \ }}}
\put(55,135){\parbox{7cm}{\large \bf \ \ \ \ \ bound entangled states }}
\put(205,93){\oval(80,60)}
\put(173,93){\parbox{2.6cm}{\large \bf \ \ \ \ \ free\\  
\phantom{l}entangled\\ \centerline{states \ }}}
%
%
\put(150,105){\line(1,1){15}}
\put(150,90){\line(1,1){15}}
\put(150,75){\line(1,1){15}}
\put(150,60){\line(1,1){16}}
\put(150,45){\line(1,1){22.3}}
\put(161,40){\line(1,1){22.3}}
\put(176,40){\line(1,1){22.5}}
\put(191,40){\line(1,1){22.5}}
\put(206,40){\line(1,1){22.5}}
\put(221,40){\line(1,1){49}}
\put(236,40){\line(1,1){34}}
\put(251,40){\line(1,1){19}}
\put(245,79){\line(1,1){25}}
\put(245,94){\line(1,1){25}}
\put(244,109){\line(1,1){26}}
\put(237.5,118.5){\line(1,1){30}}
\put(225,123){\line(1,1){34}}
\put(228,143){\line(1,1){17}}
\put(216,148){\line(1,1){12}}
\put(204,150){\line(1,1){10}}
\put(189,150){\line(1,1){10}}
\put(174,150){\line(1,1){10}}
\put(159,150){\line(1,1){10}}
\put(147,153){\line(1,1){7}}
\end{picture}}
\caption[Entanglement and distillability of mixed states]{Entanglement
and distillability of mixed states for $2\otimes 2$ and $2\otimes3$ system
\index{system!$2\otimes 2$}\index{system!$2\otimes 3$}
(a) and for higher dimensions (b). The slanted pattern denotes the
hypothetical set of bound entangled NPT states}
\label{fig-dist}
\end{figure}
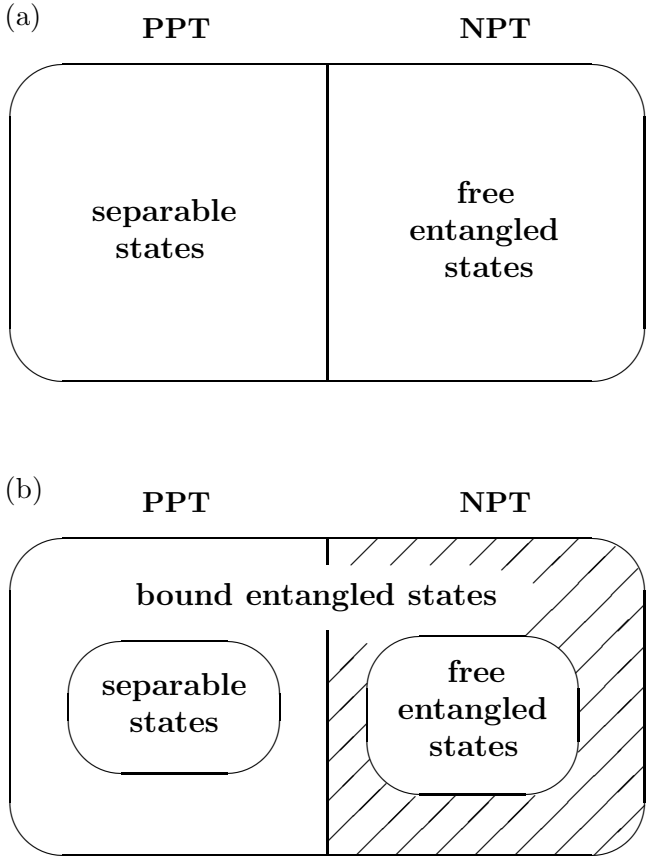

The above proposition implies that to determine if there exist
NPT bound entangled states\index{state!NPT bound entangled}, one can
restrict to the family of Werner states\index{state!Werner}
which is one parameter family of very high symmetry. Even after such a
reduction of the problem, the latter remains extremely difficult.
In  \cite{LewNPT,DiVinNPT} the authors examine the $n$-th tensor power of
Werner states\index{state!Werner} (in \cite{DiVinNPT} a larger,
two-parameter
family is considered). The results, though not conclusive yet, strongly suggest
that there
exist NPT bound entangled states\index{state!NPT bound entangled}
(see Fig. \ref{fig-dist}).

 Thus it is likely that the characterisation
of distillable states\index{state!distillable} is not  so
simple to reduce to NPT
condition\index{condition!NPT}.
Possible existence of the NPT bound
entanglement\index{entanglement!NPT bound} would make the total
picture much more obscure (hence much more interesting). Among others, there
would arise a question: for two distinct BE
states\index{state!bound entangled (BE)} $\varrho_1$ and
$\varrho_2$ is the state\index{state} $\varrho_1\otimes \varrho_2$ also BE? (if BE would
equal PPT, this question has immediate answer ``yes'', because the PPT property
is additive, i.e. if two states\index{state!PPT} are PPT, then
so their
tensor product\index{product!tensor} does
\cite{Peres}). Recently, negative answer to this question was 
obtained in Ref. \cite{superactivation} in the case of 
multipartite
system\index{system!multipartite}. For bipartite 
states\index{state!bipartite} the answer is still unknown.

\section{Example}
\label{sec-bound-examples}
Consider the family of states\index{state} (\ref{eq-Stormer}) considered
in Sect.
\ref{examples}. One obtains the following classification: $\varrho$ is
\bei
\item separable for $2\leq\alpha\leq3$
\item BE for $3<\alpha\leq4$
\item FE for $4<\alpha\leq5$
\eei
Separability was shown in Sect. (\ref{examples}). It was also shown there
that for $3<\alpha\leq4$ the state is entangled and
PPT\index{state!PPT entangled}. Then we conclude that
it is BE. For $\alpha>4$ Alice and Bob can apply
local projectors\index{projector!local}
$P=|0\ra\la0|+|1\ra\la1|$  obtaining  entangled two-qubit state%
\index{state!entangled}\index{state!two-qubit}. Hence the
initial state\index{state!free entangled (FE)} is FE in this region of $\alpha$.

\section{Some consequences of existence of bound entanglement}
\label{sec-conseq}

A basic question that arises in the context of
bound entanglement\index{entanglement!bound} is:
What is its role
in the quantum information\index{quantum!information}%
\index{information!quantum} theory? We will show in next sections,
that despite it is indeed a very poor type
of entanglement\index{entanglement!types}, it can produce
non-classical effect, enhancing
quantum communication\index{quantum!communication}%
\index{communication!quantum} via a subtle
activation-like\index{activation} process \cite{aktyw}. This will lead us to a new paradigm of
entanglement processing\index{entanglement!processing}
extending the ``LQCC paradigm''. Moreover,
the existence of bound entanglement\index{entanglement!bound} means
that there exist stronger
limits for distillation rate\index{distillation!rate} than it
was expected before.
These, and other consequences we will report in the next few subsections.

\subsection{Bound entanglement and teleportation}
By definition, BE states\index{state!bound entangled (BE)} cannot be
distilled, hence it is impossible to
obtain faithful teleportation\index{teleportation!faithful}
via such state\index{state}.
However, it might be the case
that the transmission\index{transmission!fidelity}
fidelity\index{fidelity!transmission} of {\it imperfect}
teleportation\index{teleportation!imperfect}
is still
better than the one achievable by purely classical
channel\index{channel!classical} i.e. without
sharing any entangle\-ment\index{entanglement} (this was a way of
revealing manifestation of quantum features of some
mixed state\index{state!mixed}
\cite{Popescu94})\footnote{For detailed study of standard teleportation scheme
via mixed two-qubit state see \cite{tel}. The optimal one-way teleportation 
via pure states was obtained in \cite{Banaszek}.}. First searches 
produced a negative result
\cite{LindPop}. Here we present more general  results, according to which
the most  general teleportation
scheme\index{teleportation!scheme!general}
 cannot produce better than classical
fidelity\index{fidelity!classical}, if Alice and Bob share
BE states\index{state!bound entangled (BE)}.

\subsubsection{General teleportation scheme}
Teleportation\index{teleportation}, as originally
devised \cite{Bennett_tel}, is a way of
transmission\index{transmission!of quantum state} of a
quantum state\index{state!quantum}\index{quantum!state} by use
of classical
channel\index{channel!classical} and bipartite
entangled state\index{state!mixed}\index{state!bipartite}
(pure singlet state\index{state!pure}\index{state!singlet})
shared by Alice and Bob. The most general scheme of teleportation,
\index{teleportation!scheme!general} would be
then of the following form \cite{single}.  There are three
systems\index{system}: the
one of the input
particle, the state\index{state} of which is to be teleported (ascribe to this
system\index{system} the
Hilbert space\index{space!Hilbert} $\hcal_{A'})$, and two
systems\index{system},
that are in the entangled state\index{state!entangled}
$\varrho_{AB}$ (with Hilbert
space\index{space!Hilbert} $\hcal=\hcal_A\otimes\hcal_B$). For
simplicity we assume that $\dim\hcal_{A'}=\dim\hcal_{A}=\dim \hcal_{B}=d$. The
initial state\index{state} is
\[
|\psi_{A'}\ra\la\psi_{A'}|\otimes \varrho_{AB}
\]
where $\psi_{A'}$ is the state\index{state} to be teleported (unknown to Alice and Bob).
Now Alice and Bob perform some trace preserving
LQCC operation\index{operations!LQCC!trace-preserving} (trace
preserving, because teleportation\index{teleportation}, is the
operation\index{operation} that must be performed with
probability $1$). The form of the operation\index{operation} depends
on the state\index{state} $\varrho_{AB}$
that is known to Alice and Bob, but are independent of the
input state\index{state!input}
$\psi_{A'}$ because it is unknown.
Now the total system\index{system} is in new, perhaps very  complicated
state\index{state}
$\varrho_{A'AB}$. The transmitted state\index{state} is given by
$\trace_{A'A}(\varrho_{A'AB})$. The overall
transmission\index{transmission} stages are the
following
\[
\kern-2mm\psi_{A'} \rightarrow |\psi_{A'}\ra\la\psi_{A'}|\otimes \varrho_{AB}\rightarrow
\Lambda(|\psi_{A'}\ra\la\psi_{A'}|\otimes \varrho_{AB})\rightarrow
\trace_{A'A}\varrho_{A'AB}=\varrho_B\;.
\]
Now the {\it transmission
fidelity}\index{transmission!fidelity}\index{fidelity!transmission} is
defined by
\[
f=\overline{\la\psi_{A'}|\varrho_B|\psi_{A'}\ra}\;,
\]
where the average is taken over uniform distribution of the input
states\index{state!input}
$\psi_{A'}$~\footnote{Note that so defined
fidelity\index{fidelity} is not a
unique criterion of
performance of teleportation\index{teleportation}. For example,
one can consider restricted input:
Alice receives one of {\it two} nonorthogonal vectors\index{vector} 
with some probabilities
\cite{Henderson}.
Then the formula for fidelity\index{fidelity} would be different. 
In general,
fidelity\index{fidelity} is
determined by a chosen distribution over input
states\index{state!input}.}.
In the original teleportation\index{teleportation!scheme!original}
scheme (where $\varrho_{AB}$ is a maximally
entangled state\index{state!entangled!maximally}), the
state $\varrho_B$ is exactly equal to the
input state\index{state!input}, so that $f=1$.
If Alice and Bob share a pair in separable
state\index{state!separable} (or, equivalently,
share {\it no} pair), 
then the best one can do is the following: Alice measures the
state\index{state}
and sends the results to Bob \cite{Popescu94}. Since it is impossible to find
the form of the
state\index{state} having only a single system\index{system} in that state\index{state}
\cite{dAriano} (it would contradict no-cloning
theorem\index{theorem!no-cloning} \cite{Zurek})
the performance of such
process will be very poor.  One can check that the best
possible fidelity\index{fidelity} is
$f=2/(d+1)$. If the shared pair is  entangled, but it is not a 
pure maximally
entangled state\index{state!entangled!maximally}, we will obtain some
intermediate values of $f$.

\subsubsection{Optimal teleportation}
Having defined the general teleportation
scheme\index{teleportation!scheme!general},, one can ask about the maximal
fidelity\index{fidelity} that can be achieved for given
state\index{state} $\varrho_{AB}$ within the scheme. Thus, for given
$\varrho_{AB}$ we must
maximise $f$ over all possible trace-preserving
LQCC operations\index{operations!LQCC!trace-preserving}. The
problem is, in general,
extremely difficult. However the high symmetry of the chosen
fidelity function\index{fidelity!function}
allows to reduce it in the following way. It has been shown
\cite{single} that the best Alice and Bob can do is the following. They first
 perform some LQCC action\index{action!LQCC} that aims
 at increasing $F(\varrho_{AB})$
 as much as possible. Then they perform the standard
 teleportation scheme\index{teleportation!scheme!standard},
 via the new state\index{state} $\varrho'_{AB}$ (just as if it were the
 state\index{state} $P_+$).
 The obtained fidelity\index{fidelity} is given  by
 \be
f_{\max} = {F_{\max} d+1\over d+1}\;,
\label{eqtel}
\ee
where $F_{\max}=F(\varrho'_{AB})$ is the maximal $F$ that can be obtained
by trace-preserving LQCC actions\index{action!LQCC!trace-preserving} if
the initial state\index{state} is  $\varrho_{AB}$.
\subsubsection{Teleportation via bound entangled
states\index{state!bound entangled (BE)}}
According to (\ref{eqtel}), to check the performance of
teleportation\index{teleportation!via BE states}
via BE states\index{state!bound entangled (BE)} of $d\otimes d$
system\index{system!$d\otimes d$}, we should find maximal $F$
attainable from BE states\index{state!bound entangled (BE)} via trace-preserving LQCC
actions\index{action!LQCC!trace-preserving}. As it was argued in Sect.
\ref{sec-bound} a BE state\index{state!bound entangled (BE)} subjected to any
LQCC operation\index{operation!LQCC} remains BE.
Moreover singlet fraction\index{fraction!singlet} $F$ of
a BE states\index{state!bound entangled (BE)} of $d\otimes d$
system\index{system!$d\otimes d$} satisfies $F\leq1/d$ (because
states\index{state!distillable} with $F>1/d$ are
distillable, as shown in Sect. \ref{sec-dist-examples}).
 We conclude that
if the initial state is BE\index{state!bound entangled (BE)} then the
highest $F$ achievable by any
(not only trace preserving) LQCC actions\index{action!LQCC}
is $F=1/d$. However, as we have argued, this gives
fidelity\index{fidelity}
$f=2/(d+1)$, that
can be achieved without entanglement\index{entanglement}, too. Thus
the BE states\index{state!bound entangled (BE)} behave here
like separable states\index{state!separable} -- their
entanglement\index{entanglement} does
not manifest itself.

\subsection{Activation of bound entanglement}
\label{aktywacja}
Here we will show that bound entanglement\index{entanglement}
can produce a non-classical effect,
 even though  the effect is  very subtle one. It is the so-called 
{\it activation of bound entanglement} \cite{aktyw}. 
The underlying concept originates
from formal 
entanglement-energy analogy\index{entanglement-energy analogy} developed in
\cite{term1,term,VP98,clon,bound}. One can
imagine that the
bound entanglement\index{entanglement!bound} is like energy\index{energy} of the
system\index{system} confined in a shallow
potential well. Then, as in the process of
chemical activation,
if we add
a small amount of extra energy\index{energy} to the
system\index{system}, its
energy\index{energy} can be liberated.

In our case, the role of the system\index{system} will be
played by a huge amount of
bound entangled pairs, while the extra
energy\index{energy} -- by a {\it single} pair that
is free entangled. More specifically, we will show that a process called {\it
conclusive teleportation}\index{teleportation!conclusive} \cite{Tal}
can be performed with arbitrarily
high fidelity\index{fidelity} if Alice and Bob can perform
joint operations\index{operations} over the
BE pairs and the FE pair. We will argue that it is impossible if
either of the two elements is lacking.

\subsubsection{Conclusive teleportation}

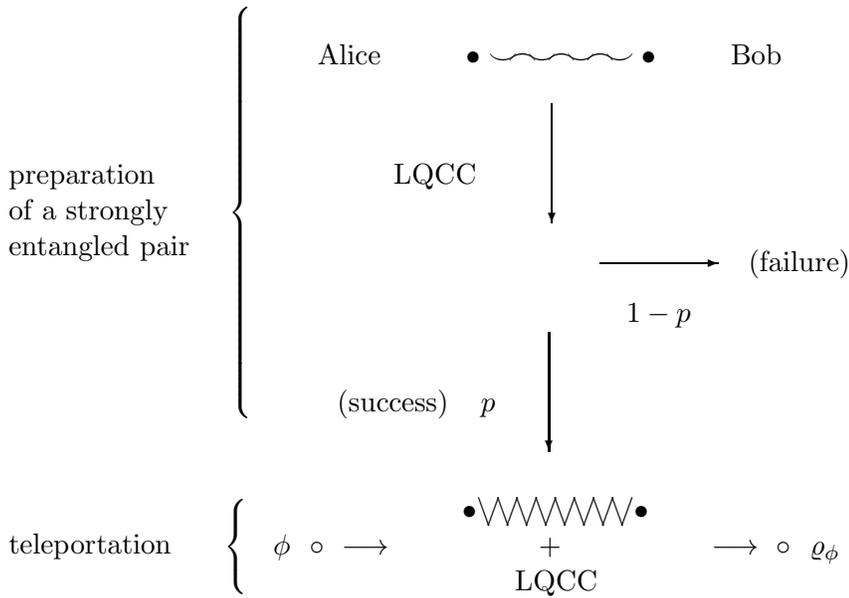
\begin{figure}
\begin{minipage}{1.1in}preparation

of a strongly

entangled pair
\end{minipage}
$\left\{\hskip0cm\parbox{3in}{
\vskip5mm
\centerline{\hfill \hskip3mm Alice \hskip1cm
$\bullet \smile\!\!\frown\!\!\smile\!\!\frown
\!\!\smile\!\!\frown\!\!\smile
\bullet$\hskip 1cm  Bob\hfill}
\vskip5mm
\centerline{LQCC
\begin{picture}(40,30)
\put(25,30){\vector(0,-1){45}}
\end{picture}
\phantom{LQCC}}
\vskip-8mm
\centerline{\hskip2.5cm
\begin{picture}(80,30)
\put(70,-20){\vector(1,0){45}}
\end{picture}
\hskip1.5cm
\lower0.8cm\hbox{(failure)}}
\centerline{\phantom{$1$
\begin{picture}(80,30)
\put(70,-20){\vector(1,0){45}}
\end{picture}
} \hskip-2mm\raise5mm\hbox{$1-p$}}
\vskip-4mm
\centerline{(success) \hskip3mm $p$ \kern-1cm
\begin{picture}(80,30)
\put(45,30){\vector(0,-1){45}}
\end{picture}
\phantom{success}}}
\right.$
\vskip5mm\vskip5mm

$\parbox{1.1in}{teleportation}
\left\{\hskip0cm\parbox{3in}{
\centerline{\hskip6mm
$\bullet\backslash\!/\!\backslash\!/\!\backslash\!/\!\backslash\!/%
\!\backslash\!/ 
\!\backslash\!/\!\backslash\!/\!%
\backslash\!/\bullet$}
\centerline{\hskip6mm
$\phi\ \  \circ \ \longrightarrow $\hskip2cm $+$\hskip2cm $
\longrightarrow \ \circ
\ \ \varrho_\phi$}
\centerline{\hskip6mmLQCC}    }\right.$
\vskip5mm
\caption[Conclusive teleportation]{Conclusive
teleportation. Starting
with a weakly entangled pair Alice and Bob prepare with probability $p$
a strongly entangled pair and then perform teleportation}
\label{fig-conc-tel}
\end{figure}

Suppose that Alice and Bob have a pair in a
state\index{state} for which the optimal
teleportation\index{teleportation!fidelity}
fidelity\index{fidelity!of teleportation, optimal}
is $f_0$. Suppose
further, that the fidelity\index{fidelity} is too
poor for some Alice and Bob purposes. What they can  do to change the situation
is to perform the so-called conclusive
teleportation\index{teleportation!conclusive}. Namely, they
can perform some LQCC operation\index{operation!LQCC} with
two final outcomes 0 and 1. Obtained the
outcome 0 they fail and decide to discard the pair. If the outcome is 1 they
perform
teleportation\index{teleportation}, and the fidelity\index{fidelity} is
now better than the initial
$f_0$. Of course, the price they must pay is that the probability of the
success (outcome 1) may be small. The scheme is illustrated on Fig.
\ref{fig-conc-tel}.

A simple example is the following.
Suppose that Alice and Bob share a pair in pure
state\index{state!pure}
$\psi=a|00\rangle+b|11\rangle$ which is nearly product
(e.g. $a$ is close to 1). Then the standard
teleportation\index{teleportation!scheme!standard} scheme  provides
a rather  poor fidelity\index{fidelity} $f=2(1+ab)/3 $ 
\cite{Gisin_tel,tel}. However, Alice can subject her particle to
filtering\index{filtering}
procedure \cite{Gisin,conc} described by the
operation\index{operation}
\begin{equation}
\Lambda=W(\cdot)W^\dagger +V(\cdot)V^\dagger
\end{equation}
with $W={\rm diag}(b,a)$, $V={\rm diag}(a,b)$. Here the outcome 1 (success)
corresponds to operator\index{operator} $W$. Indeed, if
this outcome was obtained, the state\index{state}
collapses to the singlet one
\begin{equation}
\tilde\psi={W\otimes \id\psi\over ||W\otimes
\id\psi||}={1\over\sqrt2}(|00\rangle+|11\rangle)\;.
\end{equation}
Then, in this case  perfect
teleportation\index{teleportation!perfect} can be performed. Thus, if Alice and
Bob teleported directly via the
initial state\index{state}, they would obtain
a very poor performance. Now, they have a small, but nonzero chance of
performing perfect teleportation\index{teleportation!perfect}.

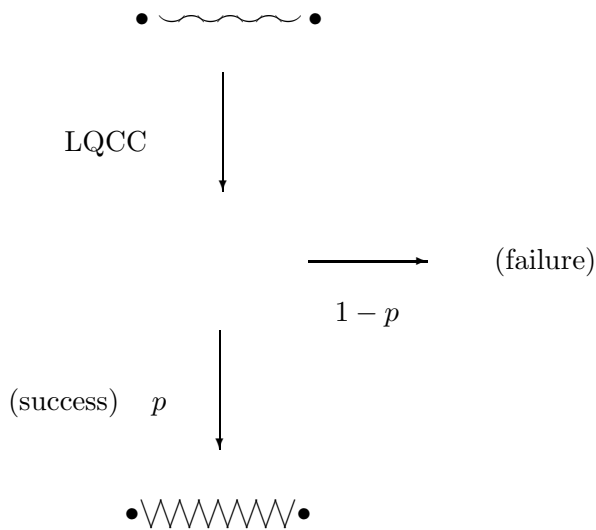
\begin{figure}
\parbox{3in}{
\hbox to \hsize{\phantom{$\varrho_{in}$} \hfill  \hskip2cm \ \ \
\hskip-2mm$\bullet \smile\!\!\frown\!\!\smile\!\!\frown
\!\!\smile\!\!\frown\!\!\smile\!\!
\hskip1mm\bullet$\hskip 1.8cm
\hfill\phantom{$\varrho_{in}$}}
\vskip5mm
\centerline{{LQCC}
\begin{picture}(40,30)
\put(25,30){\vector(0,-1){45}}
\end{picture}
\phantom{LQCC}}
\vskip-4mm
\centerline{\hskip4cm
\begin{picture}(80,30)
\put(70,-20){\vector(1,0){45}}
\end{picture}
\hskip2cm
\lower0.8cm\hbox{(failure)}}
\centerline{\phantom{$1$
\begin{picture}(80,30)
\put(70,-20){\vector(1,0){45}}
\end{picture}
} \hskip8mm\raise5mm\hbox{$1-p$}}
\vskip-4mm
\centerline{(success) \hskip3mm $p$ \kern-1cm
\begin{picture}(80,30)
\put(45,30){\vector(0,-1){45}}
\end{picture}
\phantom{success}}
\vskip5mm \vskip5mm
\centerline{\	\
$\bullet\backslash\!/\!\backslash\!/\!\backslash\!/\!\backslash\!/%
\!\backslash\!/
\!\backslash\!/\!\backslash\!/\!%
\backslash\!/\bullet$}
}
\vskip5mm
\caption[Conclusive increasing singlet fraction]
{Conclusive increasing of singlet fraction. Alice and Bob with probability
$p$ of
success obtain a state\index{state} with higher singlet fraction than the one of the
initial state\index{state}}
\label{fig-concF}
\end{figure}
Similarly as in the usual
teleportation\index{teleportation!perfect}, the conclusive
teleportation\index{teleportation!conclusive} can
be reduced to conclusive increasing $F$ 
(illustrated on Fig. \ref{fig-concF})
followed by the original
teleportation\index{teleportation!protocol}
protocol\index{protocol!teleportation}.
If in the first stage Alice and Bob obtain a state\index{state} with some $F$,
then the second stage will produce the corresponding
fidelity\index{fidelity}
$f=(Fd+1)/(d+1)$.
Thus we can restrict  our consideration to conclusive increasing singlet
fraction\index{fraction!singlet!conclusive increasing}. The latter
was developed in \cite{Kent,single}. An interesting
peculiarity of conclusive
increasing singlet fraction\index{fraction!singlet!conclusive increasing} is that sometimes it
is impossible to obtain $F=1$, but still $F$  arbitrarily close to 1 can be
obtained. However, if $F\rightarrow 1$ then the probability of success
tends to $0$, so that, indeed, it is impossible to reach $F=1$
\cite{single}.


\subsubsection{Activation protocol}
Suppose that Alice and Bob share a {\it single}
pair of spin-1\index{spin} particles in the following free entangled
mixed state\index{state!mixed}\index{state!free entangled (FE)}
\begin{equation}
\varrho_{free}=\varrho(F)\equiv F|\psi_{+}\rangle \langle \psi_{+}| +
(1-F)\sigma_{+}, \ \ 0<F<1\;,
\label{source}
\end{equation}
where $\sigma_\pm$ are separable states\index{state!separable}
given by (\ref{mix-}).
It is easy to see that the
state\index{state!free entangled (FE)} (\ref{source}) is free
entangled. Namely after action\index{action} of the
local projections $(|0\rangle \langle 0| +|1\rangle\langle 1|)\otimes
(|0\rangle\langle 0| +|1\rangle \langle 1|)$
we get an entangled $2 \otimes 2$
state\index{state!entangled}\index{state!$2\otimes2$} (its
entanglement\index{entanglement} can be revealed
by calculating partial transposition). Thus, according to the Theorem
\ref{dist-char}, the 
state\index{state!free entangled (FE)} (\ref{source}) is FE.
By  complicated considerations one can show \cite{single} that there is
a threshold $F_0<1$
that cannot be exceeded in the process of conclusive increasing
singlet fraction\index{fraction!singlet!conclusive increasing}.
In other words Alice and Bob have no chance to obtain
a state\index{state} $\varrho'$ with $F(\varrho')>F_0$
(we do not know the value $F_0$, we only know that such a number
exists).

Suppose now that Alice and Bob share in addition a very large number of pairs
in the following BE state\index{state!bound entangled (BE)} (the one considered in
Sect. (\ref{sec-bound-examples}))
\be
\sigma_{\alpha}=\frac{2}{7}|\psi_{+}\rangle \langle \psi_{+}|
+\frac{\alpha}{7} \sigma_{+}+
\frac{5-\alpha}{7} \sigma_{-}\;.
\label{target}
\end{equation}
As stated in Sect. \ref{sec-bound-examples} for $3<\alpha\leq 4$
the state is BE\index{state!bound entangled (BE)}.
As one knows, from the BE pairs of $3\otimes 3$
system\index{system!$3\otimes 3$} there is no
chance to obtain even a pair with $F>1/3$.
Now it turns out that if Alice and Bob 
have {\it both} FE pair and the BE pairs, they can apply
a simple protocol\index{protocol}  to
obtain  $F$ {\it arbitrarily close} to 1.
Thus, due to the connection between 
conclusive increasing
singlet fraction\index{fraction!singlet!conclusive increasing}
and conclusive teleportation\index{teleportation!conclusive},
the fidelity\index{fidelity} of the
latter can be arbitrarily close to unity only if
both FE pair and BE pairs  are shared.

The  protocol\index{protocol} \cite{aktyw} is similar to the
recurrence distillation\index{protocol!recurrence, distillation}
protocol\index{distillation!protocol!recurrence} described in
Sect. \ref{sec-Bennett}. It is an iteration of the following two steps
\bei
\item[(i)] \ Alice and Bob take the free entangled pair in
the state\index{state} $\varrho_{free}(F)$
and one of the pairs being in the state\index{state} $\sigma_{\alpha}$.
They perform the bilateral XOR
operation\index{operation!bilateral XOR}
$U_{BXOR}\equiv U_{XOR}\otimes U_{XOR}$, each of them treating the member
of free (bound) entangled pair as a source (target)%
\footnote{Here we need the quantum XOR\index{quantum!gate!XOR}
gate\index{gate!quantum XOR} not for two
qubits\index{qubit} as in Sect.
\ref{sec-Bennett} but for two
qutrits\index{qutrit} (three-level
systems\index{system!three-level}). A general
XOR operation\index{operation!general XOR} for $d \otimes d$ system%
\index{system!$d\otimes d$} that was used in
in \cite{xor,gott} is defined as
\begin{equation}
U_{XOR} | a \rangle | b \rangle= | a \rangle | (b + a) 
{\rm mod}\, d \rangle\;,
\end{equation}
where initial state\index{state} $| a \rangle$  ($| b \rangle$)
corresponds to source (target) state\index{state}.}.
\item [(ii)]\  Alice and Bob measure
the members of source pair in basis\index{basis} $|0\ra,|1\ra,|2\ra$.
Then they compare their results via
classical communication\index{communication!classical}.
If the compared results differ from each other
they have to discard both pairs and
then the trial of improvement of $F$ fails.
If the results agree then the trial succeeds
and they discard only the target pair,
coming back with (as we shall see) improved
source pair to the first step (i).
\eei
After some algebra one can see that the success in the step (ii)
occurs with nonzero probability
\begin{equation}
P_{F \rightarrow F'} =\frac{2F + (1-F)(5 - \alpha) }{7} \ \
\end{equation}
leading then to the
transformation\index{transformation} $\varrho(F) \rightarrow \varrho(F')$
with the improved fidelity\index{fidelity}
\begin{equation}
F'(F)=
\frac{2F}{2F + (1-F)(5-\alpha)}\;.
\label{func}
\end{equation}

If only $\alpha>3$, then  the above continuous function
of F exceeds the value of F on the whole region $(0,1)$.
Thus the successful repeating of the steps (i-ii) produces
the sequence of source fidelities $F_{n} \rightarrow 1 $.
On Fig. \ref{fig-aktyw}
\begin{figure}
\caption[Activation of bound entanglement]{Liberating bound entanglement.
The singlet fraction\index{fraction!singlet} of the
FE state\index{state!free entangled (FE)} is plotted versus
the number of successful iterations (i-ii) and the parameter $\alpha$ of the
state\index{state} $\varrho_\alpha$ of the used BE pairs. The initial singlet
fraction\index{fraction!singlet}
of the FE pair is taken $F_{in}=0.3$ (This figure is reproduced from
Phys. Rev. Lett. {\bf 82}, 1056 (1999) by permission of authors.)}
\label{fig-aktyw}
\end{figure}
we plotted the obtained $F$ versus
number of iteration of the protocol\index{protocol}
 and the parameter $\alpha$. For
$\alpha\leq3$ the singlet fraction\index{fraction!singlet} goes down:
separable states\index{state!separable} cannot
help to increase it. We can see the dramatic qualitative change
at the ``critical''\footnote{The term ``critical'' we used here
reflects the rapid character of the change (see \cite{Lew_San} for a similar
``phase transition'' between separable and FE
states\index{state!separable}\index{state!free entangled (FE)}). On they
other hand, the present
development of thermodynamical
analogies\index{thermodynamical analogy} in entanglement
processing\index{entanglement!processing}
\cite{term1,term,Plenio,clon,bound} allows to
hope that in future one will be able to build a synthetic theory
of entanglement\index{entanglement}
based on thermodynamical
analogies\index{thermodynamical analogy}: then the ``critical'' point would become
truly critical.} point  that occurs at  the borderline between separable
states\index{state!separable} and bound
entangled ones ($\alpha=3$). On the other hand it is surprising that
there is no qualitative difference between the behaviour of
BE states\index{state!bound entangled (BE)}
($3<\alpha\leq4$) and FE
states\index{state!free entangled (FE)} ($4<\alpha\leq5$).
Here the change is only
quantitative while the shape of the corresponding curves is basically the same.
To authors knowledge this is the only effect we know about, where the bound
entanglement\index{entanglement!bound} manifests its 
quantumness\footnote{In multipartite case two other effects
have been recently found \cite{superactivation,BEconc}.}.
Since the effect is very subtle,
one must conclude that bound entanglement\index{entanglement!bound} is
essentially
different from the free entanglement\index{entanglement!free}, and it
is enormously weak. For recent results on activation effect
\index{activation} in multiparticle
case see \cite{Cirac-akt}. 

\subsection{Entanglement enhanced LQCC operations}

The activation\index{activation} effect suggests to
extend the paradigm of LQCC operations\index{operations!LQCC} by
including
quantum communication\index{communication!quantum}%
\index{quantum!communication}
(under suitable control). Then we obtain {\it entanglement enhanced}
LQCC (LQCC+EE) operations\index{operations!LQCC!entanglement enhanced}
(see \cite{Lo} in this context). For example,
if we allow LQCC operation\index{operations!LQCC}
and arbitrary amount of shared bound
entanglement\index{entanglement!bound}, we obtain LQCC+BE
paradigm.  One can now ask about entanglement of
formation\index{entanglement!of formation}%
\footnote{Entanglement of
formation\index{entanglement!of formation} $E_{LQCC}^F(\varrho)$ of
a state\index{state} $\varrho$ is the amount of
input singlet pairs per output pair needed to produce
the state\index{state} $\varrho$ by LQCC operations\index{operations!LQCC}
\cite{huge}.} and distillation\index{distillation} in
this regime. Since the BE states\index{state!bound entangled (BE)} contain
entanglement\index{entanglement}, even though very weak,
then infinite amount of it could make $D_{LQCC+BE}$ much larger
than  usual $D_{LQCC}$: one might expect $D_{LQCC+BE}$ to be
maximal possible, independently of the input
state\index{state!input} $ \varrho$
\cite{bechan} (e.g. for two-qubit pairs we would have $D_{LQCC+BE}=1$ for any
state\index{state}). In \cite{UPB2,Vedral} it was shown that it is impossible.
The argument of \cite{UPB2} is as follows. First, the authors
recall that $D_{LQCC}\leq E^F_{LQCC}$ \cite{huge}. Otherwise
it would be possible to increase
entanglement\index{entanglement} by means of LQCC
actions\index{action!LQCC}.
Indeed, suppose, that for some
state\index{state} $\varrho$ we have
$D_{LQCC}(\varrho)> E^F_{LQCC}(\varrho)$. Then Alice and Bob could take
$n$  two-qubit pairs in singlet
state\index{state!singlet}, produce $n/E^F_{LQCC}$
pairs of the state\index{state} $\varrho$. Then they could distill $n(D_{LQCC}/E^F_{LQCC})$
singlets, that would be greater number than $n$.
A similar argument is applied to $LQCC+BE$
action\index{action!LQCC+BE}: the authors  show that
it is impossible to increase number of singlet pairs
by LQCC+BE actions\index{action!LQCC+BE} and
conclude that $D_{LQCC+BE}\leq E_{LQCC+BE}^F$.
On the other hand, obviously  we have $E^F_{LQCC+BE}\leq E_{LQCC}^F$.
Combining the inequalities we obtain that $D_{LQCC+BE}$ is bounded by
the usual entanglement of
formation\index{entanglement!of formation} $E^F_{LQCC}$, that is maximal
only for singlet type states\index{state}. A different  argument  
in  \cite{Vedral} bases
on Rains results on bounds for distillation of
entanglement\index{entanglement!distillation} \cite{Rains}
(see Sect. \ref{Limits}). Thus, even though employing infinite amount of 
BE pairs the LQCC+BE operations\index{operations!LQCC+BE} are
not enormously powerful. However, it is
still possible, that they are better than LQCC themselves, i.e.
one conjectures that $D_{LQCC+BE}(\varrho)>D_{LQCC}(\varrho)$ for some 
states\index{state} $\varrho$.

\subsection{Bounds for entanglement of distillation}
\label{Limits}
Bound entanglement\index{entanglement!bound} is an achievement
in qualitative description, however,
as we could see in previous section it has an impact on quantitative approach.
Here we will see, that it helped to obtain a strong upper bound for
 entanglement of distillation\index{entanglement!of distillation} $D$
 (recall that the latter has the meaning of the
capacity of the noisy teleportation\index{teleportation!channel}%
\index{teleportation!channel},
channel\index{channel!teleportation}\index{channel!noisy}
constituted by bipartite mixed
states\index{state!mixed}\index{state!bipartite}, hence is
a central parameter of quantum
communication\index{quantum!communication}\index{communication!quantum}
theory).

The first upper bound for $D$ was entanglement
of formation\index{entanglement!of formation}
\cite{huge} calculated explicitly for two-qubit
states\index{state!two-qubit} \cite{Wooters}.
However, a stronger bound has been provided in \cite{VP98} (see also
\cite{Rains1}). It is given by the following measure of entanglement%
\index{entanglement!measure}
\cite{Plenio,VP98} based
on the relative entropy\index{entropy!relative}
\be
E_{VP}(\varrho)=\inf_\sigma S(\varrho|\sigma)\;,
\label{eq-Plenio}
\ee
where  infimum is taken over all separable
states\index{state!separable} $\sigma$.
The relative entropy\index{entropy!relative} is defined by
\[
S(\varrho|\sigma)=\trace\varrho\log\varrho-\trace \varrho\log\sigma\;.
\]
Vedral and Plenio provided a tricky argumentation  \cite{VP98} showing
that $E_{VP}$ is upper bound for $D(\varrho)$,
at additional assumption, that it is  additive. Even though we still do not
know if it is indeed additive, Rains showed \cite{Rains} that it is
a bound for $D$ even without this assumption. He also obtained a stronger
bound by use of BE states\index{state!bound entangled (BE)} (more
precisely: PPT states\index{state!PPT}). It
appears, that
if the infimum in (\ref{eq-Plenio}) is taken over
PPT states\index{state!PPT} (that are
bound entangled), then the new measure $E_R$
is a bound for distillable
entanglement\index{entanglement!distillable}, too.
However, since the set of PPT
states\index{state!PPT} is strictly greater than the set of
separable states\index{state!separable}, the bound is
stronger. For example, the entangled PPT
states\index{state!PPT entangled} have zero distillable
entanglement\index{entanglement!distillable}. Since
they are not separable, $E_{VP}$ does not vanish for them,
hence the evaluation of $D$ by means of $E_{VP}$ is too rough.  The Rains
measure  vanishes for these states\index{state}.

We will not provide here the original proof of the Rains result. Instead we
demonstrate a general theorem on bounds for distillable entanglement%
\index{entanglement!distillable} obtained in \cite{miary}, that
allow essential simplification of
the proof of the result.

 \begin{theorem} \rm
 Any function $B$ satisfying the
conditions a)-c) below is an upper bound for entanglement of distillation%
\index{entanglement!distillable}:

{\parindent=0pt
a) Weak monotonicity\index{monotonicity!weak}:
$B(\varrho)\geq B(\Lambda(\varrho))$ where
$\Lambda$ is superoperator\index{superoperator}  realizable by means  of
LQCC operations\index{operations!LQCC}.
\vskip1mm
b) Partial
subadditivity\index{subadditivity}: $B(\varrho^{\otimes n})\leq nB(\varrho)$
\vskip1mm
c) Continuity for isotropic state\index{state!isotropic} $\varrho(F,d)$:
suppose that we have a sequence of isotropic states\index{state!isotropic}
 $\varrho(F_d,d)$,
(see Sect. \ref{examples}, formula (\ref{isotropic}))
such that $F_d\rightarrow 1$ if $d\rightarrow \infty$.
Then we require
\be
\lim_{d\rightarrow \infty} {1\over \log d} B(\varrho(F_d,d))\rightarrow 1\;.
\ee}
\end{theorem}

\textbf{Remarks.}
If, instead of LQCC operations\index{operations!LQCC} we
take other class $C$ of operations\index{operations}
including classical communication\index{communication!classical} at
least in one direction
(e.g. the mentioned LQCC+BE operations\index{operations!LQCC+BE}),
the proof {\it mutatis mutandis} also applies. (then the
condition a) would involve the class $C$)

\textbf{Proof.}
The main idea of the proof is to exploit the
monotonicity\index{monotonicity!condition}
condition\index{condition!monotonicity}:
We will show that if $D$ were greater than $B$ then during
distillation
protocol\index{distillation!protocol} the function $B$ would
have to increase. But it cannot be so, because
distillation\index{distillation}  is
LQCC action\index{action!LQCC}, hence $B$ would violate the assumption a).
By subadditivity\index{subadditivity} we have
\be
B(\varrho)\geq {1\over n}B(\varrho^{\otimes n})\;.
\label{dowod_ed1}
\ee
Distillation\index{distillation}  of $n$ pairs
aims at obtaining $k$ pairs each
in nearly singlet state\index{state!singlet}. Then the asymptotic
rate is $\lim k/n$.
It was shown \cite{Rains} that equally well  one can
think of final $d\otimes d$ system\index{system!$d\otimes d$}  in the
state\index{state} close to $P_+^d$. The asymptotic rate is
now $\lim (\log d)/n$. Then the only relevant parameters of the final
state\index{state}
$\varrho_{out}$ is dimension $d$ and
fidelity\index{fidelity} $F(\varrho_{out})$. Thus
distillation protocol\index{distillation!protocol}
\index{protocol!of distillation} can be
followed by $U\otimes U^*$ twirling\index{twirling!$U\otimes U^*$},
producing isotropic final
state\index{state!isotropic} $\varrho(d,F)$ (see
Sect. \ref{examples}).  By condition a),
distillation\index{distillation}  does not increase
$B$, hence
\be
{1\over n} B(\varrho^{\otimes n})\geq {1\over n} B(\varrho(F_{d_n},d_n))\;.
\ee
Now, in distillation process\index{distillation!process}
$F\rightarrow 1$, and if we consider
optimal protocol\index{protocol!optimal}, then $(\log
d)/n\rightarrow D$. Hence, by condition c) the right hand side of
the inequality tends to $D(\varrho)$. Thus we obtain
that $B(\varrho)\geq D(\varrho)$.~\blacksquare

Now we should check, if the Vedral-Plenio and Rains measures satisfy
the assumption of the theorem.
Subadditivity\index{subadditivity}, and
weak monotonicity\index{monotonicity!weak}  are
immediate consequence of the
properties of relative entropy\index{entropy!relative} used in definition of $E_r$
(subadditivity\index{subadditivity} proved in \cite{Plenio},
weak monotonicity\index{monotonicity!weak} --
in \cite{VP98}).
The calculation of $E_r$ for isotropic
state\index{state!isotropic} is a little bit more involved,
but by using high symmetry of the
state\index{state}  it was found to be
\cite{Rains} $E_{VP}(\varrho(F,d))=E_R(\varrho(F,d)) =\log d +F\log F +
(1-F)\log{1-F\over d-1}$.
Evaluating now this expression for large $d$ we easily obtain
that the condition c) is satisfied. The argumentation applies without any
change to the Rains  bound.

Finally let us note that the Rains entanglement
measure\index{entanglement!measure}
attributes no entanglement\index{entanglement} to some entangled
states\index{state!entangled} (to PPT
entangled\index{state!PPT entangled} ones). Normally we
would require that the natural
postulate for entanglement
measure\index{entanglement!measure} would be:
entanglement measure\index{entanglement!measure} should vanish
if and only if the state\index{state!separable}
is separable. However, then we would have to remove
distillable entanglement\index{entanglement!distillable}
from the set of measures. Indeed,
distillable entanglement\index{entanglement!distillable} vanishes for
manifestly entangled states\index{state!entangled} - bound
entangled ones. Now the problem is:
Should we keep the postulate, or keep $D$ as a good measure?

It is reasonable to keep $D$ as a good measure, as it has direct physical
sense: it describes entanglement\index{entanglement} as a
resource for\index{resource!for quantum communication}
quantum
communication\index{quantum!communication}\index{communication!quantum}.
If it is not a measure, then we should conclude that we  are not
interested in measures. Consequently, we adopt as a main ``postulate''
for entanglement measure\index{entanglement!measure}
the statement: ``Distillable
entanglement\index{entanglement!distillable} is a good
measure\index{entanglement!measure}''.
So we must abandon the postulate. The apparent paradox
can be removed by realizing that we have  different {\it types} 
of entanglement\index{entanglement!types}.
Then a given state\index{state} even though entangled may not contain some type
of entanglement\index{entanglement!types}, and
the measure that quantifies this type will
attribute  no entanglement\index{entanglement} to
the state\index{state}.

\part{Concluding remarks}
In contrast to pure states\index{state!pure} case the 
problem of mixed-state
entanglement\index{entanglement!mixed-state} is
``non-degenerate'' in the sense that various scalar and 
structural separability 
criteria\index{criterion!separability!structural} are not
equivalent. There is a fundamental connection between
entanglement\index{entanglement} and positive
maps\index{map!positive} represented by Theorem \ref{dod}. 
However, still there is a problem of turning it into an 
operational criterion for higher-dimensional 
systems\index{system!higher-dimensional}.
Recently \cite{Le00a,Le00b} the question was reduced to 
problem of investigation  of  the so-called
``edge'' PPT entangled states \index{states!PPT entangled}
as well as the positive maps \index{map!positive} and 
entanglement witnesses\index{entanglement!witness}
detecting their entanglement\index{entanglement}.
Some operational criteria for low rank density matrices 
(also for multiparticle case) have been worked out in \cite{Ho00a}.

It is
remarkable that structure of entanglement\index{entanglement} reveals
discontinuity. There 
are two qualitatively different types of
entanglement\index{entanglement!types}: distillable  --
``free'' entanglement\index{entanglement!free} and
the ``bound''\index{entanglement!bound}
one, that cannot be distilled.
All the two qubit entangled
states\index{state!mixed}\index{state!two-qubit} are
free entangled\index{state!free entangled (FE)}.
Moreover, free
entangled state\index{state!free entangled (FE)} in any dimension must have some features of two-qubit
entanglement\index{entanglement!two-qubit}. The bound
entanglement\index{entanglement!bound}  is practically
useless for quantum
communication\index{communication!quantum}\index{quantum!communication}.
However it is not a marginal phenomenon, as the volume
of the set BE states\index{state!bound entangled (BE)} in the set of all
states\index{state} for finite dimension is nonzero.

Activation of bipartite bound entanglement suggested \cite{aktyw} nonadditivity
of corresponding quantum communication channels\footnote{It could be then reformulated in terms of the so called {\it binding entanglement channels} 
\cite{bechan,UPB2}.} in a sense, that distillable
entanglement $D(\varrho_{BE}\otimes\varrho_{EF})$ could exceed 
$D(\varrho_{FE})$ for some free entangled $\varrho_{FE}$ and bound entangled
$\varrho_{BE}$ state. 
 Quite recently it has been shown \cite{superactivation}
that in the multipartite case two different
bound entangled states\index{state!bound entangled (BE)}, 
if tensored together, can make a distillable state: $D(\varrho_{BE}^1\otimes
\varrho_{BE}^2)>D(\varrho_{BE}^1)+D(\varrho_{BE}^2)=0$.
This new nonclassical effect was called {\it superactivation}. On the other 
hand in \cite{BEconc} it was shown that the four-party 
``unlockable''  bound entangled states\index{state!bound entangled (BE)}  
\cite{unlockable} can be used for remote 
concentration of quantum information. 
It is intriguing that for bipartite systems, with the exception of the
activation\index{activation} effect, the bound
entanglement\index{entanglement!bound} is permanently passive.
In general, there may be a qualitative difference 
between bipartite bound entanglement\index{entanglement!bound} 
and multipartite one.
Still, in the light of the recent results 
\cite{Sh00} it is quite possible that also  bipartite
bound entanglement\index{entanglement!bound} is  nonadditive.
The very recent investigations of  bound 
entanglement\index{entanglement!bound} for continuous 
variables  \cite{Ho00b,We00}  rise  analogous 
questions also in this latter  domain.

As  we have seen there is a basic connection between bound
entanglement\index{entanglement!bound} and
irreversibility\index{irreversibility}.
Then it would be interesting to investigate some dynamical features of BE.
It cannot be excluded that some systems\index{system} involving
BE states\index{state!bound entangled (BE)} may reveal
nonstandard (non-exponential) decay of
entanglement\index{entanglement}.    
In general, it seems that the role of bound
entanglement\index{entanglement!bound} in quantum\index{quantum!communication}
communication\index{communication!quantum} will be negative: in fact, existence
of
BE constitutes a fundamental {\it restriction} for entanglement
processing\index{entanglement!processing}. One
can speculate that it is ultimate restriction in the context of
distillation\index{distillation},
i.e. that it may allow to determine the value distillable
entanglement\index{entanglement!distillable}.
Then it seems important to develop the approach combining BE
and the entanglement measures\index{entanglement!measure} involving relative
entropy\index{entropy!relative}.
It also seems reasonable to conjecture that in the case of the general
distillation\index{distillation} processes involving
mixed states\index{state!mixed} conversion
$\varrho\rightarrow\varrho'$ \cite{huge} bound
entanglement\index{entanglement!bound} $E_B$
never decreases%
\footnote{Bound entanglement\index{entanglement!bound} can
be quantified \cite{bound} as the
difference between entanglement of
formation\index{entanglement!of formation} and entanglement of
distillation\index{entanglement!distillable} (defined within the original
distillation\index{distillation!scheme} scheme):
$E_B=E_F-D$.} (i.e. $\Delta E_B\geq0$) in optimal processes.

The irreversibility\index{irreversibility} inherently
connected with distillation\index{distillation}  encourages to develop
some natural  {\it formal} analogies between
mixed-state entanglement\index{entanglement!mixed-state} processing
and phenomenological thermodynamics\index{thermodynamics!phenomenological}.
The  construction
of ``thermodynamics of entanglement''\index{thermodynamics!of entanglement}%
(cf. \cite{term1,clon,term,balance})
would be essential for
a synthetic understanding of entanglement
processing\index{entanglement!processing}. Of course, the progress  
in the above direction would require to develop various techniques of 
search for bound entangled states\index{state!bound entangled (BE)}.

One of the challenges of mixed-state entanglement theory 
is to determine which states are useful for quantum 
communication at given additional resources. In particular,
one still does not know: (i) which states are distillable under 
LOCC (i.e. which states are free entangled) (ii) which states 
are distillable under {\it one-way} classical communication
and local operations. 

A promising direction of mixed-state entanglement%
\index{entanglement!mixed-state}
theory is application
to the theory of quantum channel\index{channel!quantum} capacity%
\index{capacity!of quantum channel}, 
pioneered in \cite{huge}.
In particular, the methods leading to upper bounds for distillable 
entanglement\index{entanglement!distillable}
in Sect. \ref{Limits}  allow to obtain upper bounds 
for quantum channel\index{channel!quantum} 
capacities\index{capacity!of quantum channel}
\cite{shannon} (one of them obtained earlier 
in Ref. \cite{BKN}). It has been shown \cite{shannon} that the 
following hypothetical inequality 
\be
D_1(\varrho)\geq S(\varrho_B)-S(\varrho)
\ee
where $D_1(\varrho)$ is {\it one-way} distillable 
entanglement\index{entanglement!distillable}\footnote{%
Classical messages can be sent only from Alice to Bob during 
distillation\index{distillation}}, would 
imply {\it equality} between capacity\index{capacity!of quantum channel} 
of quantum channel and the maximal rate of coherent 
information\index{information!coherent}
\cite{Nielsen2}. The latter equality would be nothing but quantum 
Shannon theorem\index{theorem!Shannon}, with coherent information%
\index{information!coherent} being the counterpart of 
mutual information \index{information!mutual}.
All the results obtained so far in the domain of quantifying 
entanglement\index{entanglement!quantifying}
indicate that the inequality is true. However,  the proof of the 
inequality has not been found so far.

Finally, one would like  to have a clear  connection between
entanglement\index{entanglement} and
its basic manifestation  -- nonlocality\index{nonlocality}. One can
assume  that free
entangled states\index{state!free entangled (FE)} exhibit
nonlocality\index{nonlocality} via
distillation\index{distillation} process
\cite{hidden,Bennett_pur}. However, the question concerning possible 
nonlocality\index{nonlocality} of BE
states\index{state!free entangled (FE)} remains
open (see \cite{Peres2,Werner2,Terhal2}
in this context).

To answer the above and many other questions, one must develop the
mathematical description of the structure of
mixed-state entanglement\index{entanglement!mixed-state}. In this
context, it would be especially important to
push forward the mathematics of positive
maps\index{map!positive}. One hopes that the exciting
physics connected with  by mixed-state
entanglement\index{entanglement!mixed-state} we presented in this
contribution will  stimulate the progress in this domain.





\addcontentsline{toc}{part}{Index}
\printindex

\flushbottom



\begin{thebibliography}{99}
\addcontentsline{toc}{part}{References}

\bibitem{Peres-book}
A. Peres, {\it Quantum Mechanics: Concepts and Methods}, Kluwer, Dordrecht
(1993).
\bibitem{desp}
B. d'Espagnat: {\it Conceptual Foundations of Quantum Mechanics}, Benjamin,
Reading, Massachusetts (1976).
\bibitem{Bohm}
D. Bohm: Phys. Rev. \textbf{85} 166 (1952). 
\bibitem{EPR}
A. Einstein, B. Podolsky and N. Rosen: Phys. Rev.  \textbf{47}, 777  (1935).
\bibitem{Schrodinger}
E. Schr\"odinger:  Nat\"urwissenschaften \textbf{23}, 807 (1935).
\bibitem{Heitler}
W. Heitler and F. London: Zeits. Phys. {\bf 44}, 455 (1927). 
\bibitem{Bell}
J. S. Bell: Physica (N.Y.)  \textbf{1}, 195  (1964).
\bibitem{world}
Phys. World, March 1998; J. Gruska {\it Quantum Computing}, McGraw-Hill, 
London (1999).
\bibitem{Bennett_tel}
C. Bennett, G. Brassard, C. Crepeau, R. Jozsa, A. Peres and W. K. Wootters:
Phys. Rev. Lett.  \textbf{70}, 1895  (1993); for experimental realisation see
D. Bouwmeester, J.-W. Pan, K. Mattle, M. Elbl, H. Weinfurter and
A. Zeilinger, Nature (London) \textbf{390}, 575 (1997);
D. Boschi, S. Brance, F. De Martini, L. Hardy and S. Popescu,
Phys. Rev. Lett. \textbf{80}, 1121 (1998); A. Furusawa, J. L. S\o rensen,
S. L. Braunstein, C. A. Fuchs, H. J. Kimble and E. S. Polzik:
Science \textbf{282}, 706 (1998); M. A. Nielsen, E. Knill and R. Laflamme:
Nature \textbf{396}, 52 (1998).
\bibitem{Werner}
R. F. Werner: Phys. Rev. A \textbf{40}, 4277 (1989).
\bibitem{Popescu94}
S. Popescu: Phys. Rev. Lett. \textbf{72}, 797 (1994).
\bibitem{hidden}
S. Popescu: Phys. Rev. Lett. \textbf{74}, 2619 (1995).
\bibitem{Bennett_pur}
C. H. Bennett, G. Brassard, S. Popescu, B. Schumacher, J. Smolin and
W. K. Wootters: Phys. Rev. Lett. \textbf{76}, 722 (1996).
\bibitem{bound}
M. Horodecki, P. Horodecki and R. Horodecki:
Phys. Rev. Lett. \textbf{80}, 5239 (1998).
\bibitem{sep}
M. Horodecki, P. Horodecki and R. Horodecki: Phys. Lett. A
\textbf{223}, 1 (1996).
\bibitem{Kraus}
K. Kraus: {\it States, Effects and Operations: Fundamental Notions of
Quantum Theory} (Wiley, New York, 1991).
\bibitem{huge}
C. H. Bennett, D. P. DiVincenzo, J. Smolin, and W. K. Wootters:
Phys. Rev. A \textbf{54}, 3824 (1996).
\bibitem{Plenio}
V. Vedral, M. B. Plenio, M. A. Rippin and P. L. Knight:
    Phys. Rev. Lett. \textbf{78}, 2275 (1997).
\bibitem{VP98}
V. Vedral and M. Plenio: Phys. Rev. A \textbf{57}, 1619 (1998).
\bibitem{Tarrach}
G. Vidal and R. Tarrach: Phys. Rev. A, \textbf{59}, 141 (1999).
\bibitem{Vidal}
G. Vidal: J. Mod. Opt. \textbf{47}, 355 (2000).
\bibitem{Murao}
M. Murao, M. B. Plenio, S. Popescu, V. Vedral and P.L. Knight:
Phys. Rev. A \textbf{57}, R4075 (1998); W. D\"ur, J. I. Cirac and R. Tarrach:
Phys. Rev. Lett. \textbf{83}, 3562 (1999);
N. Linden, S. Popescu and A. Sudbery:
{\it ibid.} \textbf{83}, 243 (1999).
\bibitem{UPB1}
C. H. Bennett, D. DiVincenzo, T. Mor, P. Shor, J. Smolin and
B. Terhal: Phys. Rev. Lett. \textbf{82}, 5385 (1999).
\bibitem{transp}
P. Horodecki: Phys. Lett. A \textbf{232}, 233 (1997).
\bibitem{Schmidt}
E. Schmidt: Math. Ann. \textbf{63}, 433 (1907).
\bibitem{Gisin_Bell}
N. Gisin: Phys. Lett. A  \textbf{154}, 201 (1991).
\bibitem{renyi}
R. Horodecki, P. Horodecki and M. Horodecki: Phys. Lett. A
\textbf{210}, 377  (1996); R. Horodecki and M. Horodecki:
Phys. Rev. A \textbf{54}, 1838 (1996).
\bibitem{CHSH}
J. F. Clauser, M. A. Horne, A. Shimony and R. A. Holt: Phys. Rev. Lett.
\textbf{23}, 880 (1969).
\bibitem{bell}
R. Horodecki, P. Horodecki and M. Horodecki: Phys. Lett. A
\textbf{200}, 340 (1995).
\bibitem{Schumacher}
B. Schumacher: Phys. Rev. A \textbf{51}, 2738 (1995).
\bibitem{garg}
P. Horodecki and R. Horodecki: Phys. Rev. Lett. \textbf{76}, 2196 (1996). 
\bibitem{red}
R. Horodecki and P. Horodecki: Phys. Lett. A \textbf{194}, 147 (1994).
\bibitem{xor}
M. Horodecki and P. Horodecki: Phys. Rev. A \textbf{59}, 4206 (1999).
\bibitem{PSTT}
P. Horodecki, J. A. Smolin, B.Terhal and A. V. Thapliyal: quant-ph/99100122.
\bibitem{Marek}
M. \.Zukowski, R. Horodecki, M. Horodecki and P. Horodecki:
Phys. Rev. A \textbf{58}, 1964 (1998).
\bibitem{Peres}
A. Peres:   Phys. Rev. Lett.  \textbf{77}, 1413 (1996)
\bibitem{Giedke}
L.-M. Duan, G. Giedke, J. I. Cirac and P. Zoller:
Phys. Rev. Lett. \textbf{84}, 2722 (2000). 
\bibitem{Simon}
R. Simon: Phys. Rev. Lett. \textbf{84}, 2726 (2000).
\bibitem{Choi}
M. D. Choi: Linear Algebra Appl. \textbf{12}, 95 (1975).
\bibitem{Stormer}
E. St\o{}rmer: Acta. Math. \textbf{110}, 233 (1963).
\bibitem{Woronowicz}
S. L. Woronowicz: Rep. Math. Phys. \textbf{10}, 165 (1976).
\bibitem{Terhal-phd}
B. Terhal: Quantum Algorithms and Quantum Entanglement, PhD Thesis,
ISBN 90-9013009-8, Universiteit van Amsterdam, Amsterdam (1999).
\bibitem{Jamiolkowski}
A. Jamio\l{}kowski: Rep. Math. Phys.  \textbf{3}, 275 (1972).
\bibitem{Osaki}
H. Osaka: Linear Algebra Appl. \textbf{135}, 73 (1991).
\bibitem{broad}
V. Bu\v{z}ek, V. Vedral, M. B. Plenio, P. L. Knight, and M. Hillery:
Phys. Rev. A \textbf{55}, 3327 (1997); S. Bandyopadhyay and G. Kar:
Phys. Rev. A \textbf{60}, 3296 (1999). 
\bibitem{net}
V. Bu\v{z}ek, S.L. Braunstein, M. Hillery and D. Bruss:
Phys. Rev. A \textbf{56}, 3446 (1997);
V. Bu\v{z}ek and M. Hillery: Phys. Rev. Lett. \textbf{81}, 5003 (1998).
\bibitem{disent}
S. Bandyopadhay, G. Kar and A. Roy: Phys. Lett. A, \textbf{258}, 205 (1999).
\bibitem{Poyatos}
J. F. Poyatos, J. I. Cirac, and P. Zoller: Phys. Rev. Lett.
\textbf{78}, 390 (1997).
\bibitem{volume}
K. \.Zyczkowski, P. Horodecki, A. Sanpera and M. Lewenstein:
Phys. Rev. A \textbf{58}, 883 (1998).
\bibitem{Karol}
K. \.Zyczkowski: Phys. Rev. A \textbf{60}, 3496 (1999).  
\bibitem{Sanpera}
A. Sanpera, R. Tarrach and G. Vidal:
Phys. Rev. A \textbf{58}, 826 (1998).
\bibitem{Dagmar}
D. Bruss: Phys. Rev. A \textbf{60}, 4344 (1999).   
\bibitem{Wooters}
S. Hill and W. K. Wootters: Phys. Rev. Lett. \textbf{78}, 5022 (1997);
W. K. Wooters: {\it ibid.} \textbf{80}, 2245 (1998).
\bibitem{Eisert}
J. Eisert and M. Plenio: J. Mod. Opt. \textbf{46}, 1 (1999).
\bibitem{pur}
M. Horodecki, P. Horodecki and R. Horodecki: Phys. Rev.
Lett. \textbf{78}, 574 (1997).
\bibitem{Stormer2}
E. St\o{}rmer: Proc. Amer. Math. Soc. \textbf{86}, 402 (1982).
\bibitem{UPB2}
D. P. DiVincenzo, T. Mor, P. Shor, J. A. Smolin, B. Terhal: quant-ph/9908070.
\bibitem{Peres_Bruss}
D. Bruss and A. Peres: Phys. Rev. A \textbf{61}, 030301R (2000). 
\bibitem{Lew_San}
M. Lewenstein and A. Sanpera: Phys. Rev. Lett. \textbf{80}, 2261 (1998).
\bibitem{Terhal}
B. Terhal: quant-ph/9810091.
\bibitem{cerf}
N. Cerf, C. Adami and R. M. Gingrich: Phys. Rev. \textbf{60}, 898 (1999).
\bibitem{doktor}
P. Horodecki: Conditions for quantum separability of mixed states 
and distillation of quantum entanglement.
PhD Thesis (in Polish), Politechnika 
Gda\'nska, Gda\'nsk (1999).
\bibitem{nmr}
S. L. Braunstein, C. M. Caves, R. Jozsa, N. Linden,
S. Popescu and R. Schack: Phys. Rev. Lett. \textbf{83}, 1054 (1999)
\bibitem{LindPop}
N. Linden and S. Popescu: Phys. Rev. A \textbf{59}, 137 (1999).
\bibitem{Schack}
R. Schack and C. M. Caves: J. Mod. Opt. \textbf{47}, 387 (2000). 
\bibitem{Knill}
E. Knill, R. Laflamme: Phys. Rev. Lett. \textbf{81}, 5672  (1998).
\bibitem{Shor_alg}
P. W. Shor: {\it Proceedings of the 35th Annual Symposium on the
Foundations of Computer Science} (IEEE Press, 1994), p.124
\bibitem{Slater}
P. Slater: quant-ph/9806089; quant-ph/9810026.
\bibitem{infinite}
R. Clifton and H. Halvorson:
Phys. Rev. A \textbf{61}, 012108 (2000).
\bibitem{Ekert}
D. Deutsch, A. Ekert, R. Jozsa, C. Macchiavello, S. Popescu and A. Sanpera:
Phys. Rev. Lett. \textbf{77}, 2818 (1996).
\bibitem{Enk}
S. J. van Enk, J. I. Cirac and P. Zoller: Phys. Rev. Lett. \textbf{78}, 4293
(1997); Science \textbf{279}, 205 (1998).
\bibitem{Cover}
T. M. Cover and J. A. Thomas: {\it Elements of Information
Theory} (John Wiley and Sons, N.Y. 1991).
\bibitem{Shor}
P. Shor: Phys. Rev. A, \textbf{52}, 2439 (1995).
\bibitem{Steane}
A. Steane: Phys. Rev. Lett. \textbf{77}, 793 (1996).
\bibitem{Beth}
Th. Beth and M. Grassl: Fortschr. Phys. \textbf{46}, 459 (1998)
\bibitem{Barnum}
H. Barnum, E. Knill and M. A. Nielsen: quant-ph/9809010.
\bibitem{Bennett_cap}
C. H. Bennett, D. P.   DiVincenzo and J. Smolin: Phys. Rev. Lett. \textbf{78},
3217 (1997).
\bibitem{Bruss}
D. Bruss, D. P. DiVincenzo, A. Ekert, C. A. Fuchs,  C. Macchiavello and
J. Smolin: Phys. Rev. A \textbf{57}, 2368 (1998).
\bibitem{clon}
M. Horodecki and R. Horodecki: Phys. Lett. A \textbf{244}, 473 (1998).
\bibitem{Nielsen}
M. A. Nielsen: Phys. Rev. Lett. \textbf{83}, 436 (1999); G. Vidal:
{\it ibid.} \textbf{83}, 1046 (1999); D. Jonathan and M. B. Plenio:
{\it ibid.} \textbf{83}, 1455 (1999).
\bibitem{Briegel}
W. D\"ur, H.-J. Briegel, J. I. Cirac and P. Zoller: 
Phys. Rev. A \textbf{59}, 169 (1999).
\bibitem{d1}
D. Deutsch: Proc. R. Soc. London A \textbf{425}, 73 (1989).
\bibitem{Gisin}
N. Gisin: Phys. Lett. A \textbf{210}, 151 (1996).
\bibitem{conc}
C. H. Bennett, H. J. Bernstein, S. Popescu and B. Schumacher: Phys. Rev. A
\textbf{53}, 2046 (1996).
\bibitem{Kent}
N. Linden, S. Massar and S. Popescu: Phys. Rev. Lett. {\bf 81}, 3279 (1998);
A. Kent: {\it ibid} {\bf 81}, 2839 (1998);
A. Kent, N. Linden and S. Massar: Phys. Rev. Lett. {\bf 83}, 2656 (1999).
\bibitem{Pawel} P. Horodecki: unpublished.
\bibitem{Lewen_Cir}
M. Lewenstein, J. I. Cirac and S. Karnas: quant-ph/9903012, Phys. Rev. A 2000 
(in press).
\bibitem{Dagstuhl}
M. Horodecki: presented on Dagstuhl Seminar {\it Quantum Algorithms}
(Dagstuhl, 1998).
\bibitem{Rains}
E. M. Rains: Phys. Rev. A \textbf{60}, 179 (1999).
\bibitem{Bennett99}
C. H. Bennett, D. DiVincenzo, Ch. Fuchs, T. Mor, P. Shor, J. Smolin,
E. Rains, and W. K. Wooters: Phys. Rev. A \textbf{59}, 1070 (1999).
\bibitem{Rains1}
E. Rains: quant-ph/9707002.
\bibitem{irrev} The proof of Ref.
[M. Horodecki, P. Horodecki and R. Horodecki:  Phys. 
Rev. Lett. \textbf{84}, 4260 (2000)] requires some corrections, due 
to failure of used lemma . However the result is true, the corrected proof 
can be find in  [M. Horodecki, R. Horodecki and 
P. Horodecki, in preparation]
\bibitem{LewNPT}
D. D\"ur, J. I. Cirac, M. Lewenstein and D. Bruss: quant-ph/9910022,
Phys. Rev. A 2000 (in press).
\bibitem{DiVinNPT}
D.P. DiVincenzo, P. W. Shor, J. A. Smolin, B. Terhal and A. Thapliyal:
quant-ph/9910026.
\bibitem{superactivation}
P. W. Shor, J. Smolin and A. Thapliyal: quant-ph/0005117.
\bibitem{term}
P. Horodecki, M. Horodecki, and R. Horodecki: Acta Phys. Slovaca
\textbf{48}, 141 (1998).
\bibitem{aktyw}
P. Horodecki, M. Horodecki and R. Horodecki: Phys. Rev. Lett.
\textbf{82}, 1056 (1999).
\bibitem{tel}
R. Horodecki, M. Horodecki and P. Horodecki: Phys. Lett. A
\textbf{222}, 21  (1996).
\bibitem{Banaszek}
K. Banaszek, quant-ph/0002088.
\bibitem{single}
M. Horodecki, P. Horodecki and R. Horodecki: Phys. Rev. A
\textbf{60}, 1888 (1999).
\bibitem{Henderson}
L. Henderson, L. Hardy and V. Vedral: quant-ph/9910028.
\bibitem{dAriano}
G. M. d'Ariano and H. P. Yuen: Phys. Rev. Lett. \textbf{76}, 2832 (1996).
\bibitem{Zurek}
D. Dieks: Phys. Lett. A \textbf{92}, 271 (1982);
W. K. Wooters and W. H. \.Zurek: Nature (London) \textbf{299}, 802 (1982).
\bibitem{term1}
D. Rohrlich and S. Popescu: Phys. Rev. A \textbf{56}, 3319 (1997).
\bibitem{Tal}
T. Mor: quant-ph/9608005; T. Mor and P. Horodecki: quant-ph/9906039.
\bibitem{Gisin_tel}
N. Gisin: Phys. Lett. A \textbf{210}, 157 (1996).
\bibitem{gott}
D. Gottesmann: quant-ph/9802007.
\bibitem{Lo}
H.-K. Lo and S. Popescu: Phys. Rev. Lett. \textbf{83}, 1459 (1999).
\bibitem{BEconc}
M. Murao and V. Vedral: quant-ph/0008078.
\bibitem{Cirac-akt}
W. D\"urr and I. Cirac: quant-ph/0002028.
\bibitem{bechan}
P. Horodecki, M. Horodecki and R. Horodecki: 
J. Mod. Opt. \textbf{47}, 347 (2000). 
\bibitem{Vedral}
V. Vedral: Phys. Lett. A \textbf{262}, 121 (1999). 
\bibitem{miary}
M. Horodecki, P. Horodecki and R. Horodecki: Phys. Rev. Lett. 
\textbf{84}, 2014 (2000).
\bibitem{Le00a}
M. Lewenstein, B. Kraus, J. I. Cirac, P. Horodecki: Phys. Rev. A
\textbf{62}, 052310 (2000).
\bibitem{Le00b}
M. Lewenstein, B. Kraus, P. Horodecki, J. I. Cirac: quant-ph/0005112.
\bibitem{Ho00a}
P. Horodecki, M. Lewenstein, G. Vidal and J. I. Cirac: 
Phys. Rev. A \textbf{62}, 032310 (2000).
\bibitem{Sh00}
P. W. Shor, J. Smolin and B. Terhal: quant-ph/0010054.
\bibitem{Ho00b}
P. Horodecki and M. Lewenstein: Phys. Rev. A \textbf{85}, 2657 (2000).
\bibitem{We00}
R. F. Werner and M. M. Wolf: quant-ph/0009118.
\bibitem{unlockable}
J. Smolin: quant-ph/0001001.
\bibitem{Plenio-term}
M. B. Plenio and V. Vedral: Contemp. Phys. \textbf{39}, 431 (1998).
\bibitem{balance}
P. Horodecki, M. Horodecki and R. Horodecki: quant-ph/0002021
(Phys. Rev. A, in press).
\bibitem{BKN}
H. Barnum, M. Nielsen and B. Schumacher: \textbf{57}, 4153 (1998).
\bibitem{Nielsen2}
B. Schuamcher and M. A. Nielsen: \textbf{54}, 2629 (1996).
\bibitem{shannon}
M. Horodecki, P. Horodecki and R. Horodecki: Phys. Rev. Lett. 
\textbf{85}, 433 (2000).
\bibitem{Peres2}
A. Peres: Found. Phys. \textbf{29}, 589 (1999).
\bibitem{Werner2}
R. F. Werner and M. M. Wolf: quant-ph/9910063.
\bibitem{Terhal2}
B. Terhal: quant-ph/9911057.
\end{thebibliography}
\end{document}